\documentclass[preprint]{aastex}		

\providecommand{\eg}{e.g.}
\providecommand{\etal}{et~al.}
\providecommand{\ie}{i.e.}

\received{}
\revised{}
\accepted{}

\slugcomment{Accepted by ApJ}

\shorttitle{GRB 090417B}
\shortauthors{Holland, {\etal}}

\begin{document}


\title{GRB 090417B and its Host Galaxy: A Step Towards an
  Understanding of Optically-Dark Gamma-Ray Bursts}

\author{Stephen~T.~Holland\altaffilmark{1,2,3},
        Boris~Sbarufatti\altaffilmark{4},
        Rongfeng~Shen\altaffilmark{5},
        Patricia~Schady\altaffilmark{6},
        Jay~R.~Cummings\altaffilmark{1,3,7},
        Emmanuel~Fonseca\altaffilmark{1,3,8},
        Johan~P.~U.~Fynbo\altaffilmark{9},
        P{\'a}ll~Jakobsson\altaffilmark{10},
        Elisabet~Leitet\altaffilmark{11},
        Staffan~Linn{\'e}\altaffilmark{11},
        Peter~W.~A.~Roming\altaffilmark{8},
        Martin~Still\altaffilmark{6,12}, \&
        Bing~Zhang\altaffilmark{13}
}

\altaffiltext{1}{Astrophysics Science Division, Code 660.1,
                 8800 Greenbelt Road
                 Goddard Space Flight Centre,
                 Greenbelt, MD 20771
                 USA
                 \email{Stephen.T.Holland@nasa.gov}}

\altaffiltext{2}{Universities Space Research Association
                 10211 Wincopin Circle, Suite 500
                 Columbia, MD 21044
                 USA}

\altaffiltext{3}{Centre for Research and Exploration in Space Science and Technology
                 Code 668.8
                 8800 Greenbelt Road
                 Goddard Space Flight Centre,
                 Greenbelt, MD 20771
                 USA}

\altaffiltext{4}{INAF--IASF,
                 Via Ugo La Malfa 153,
                 I-90146 Palermo, Italy}

\altaffiltext{5}{Department of Astronomy,
                 University of Texas,
                 Austin, TX 78712
                 USA}

\altaffiltext{6}{Mullard Space Science Laboratory,
                 University College London,
                 Holmbury St Mary,
                 Dorking Surrey RH5~6NT,
                 UK}

\altaffiltext{7}{Joint Center for Astrophysics,
                 University of Maryland, Baltimore County,
                 1000 Hilltop Circle,
                 Baltimore, MD 21250
                 USA}

\altaffiltext{8}{Department of Astronomy \& Astrophysics,
                 Pennsylvania State University,
                 5252 Davey Lab,
                 University Park, PA 16802
                 USA}

\altaffiltext{9}{Dark Cosmology Centre,
                 Niels Bohr Institutet,
                 K{\o}benhavns Universitet,
                 Juliane Maries Vej 30,
                 DK--2100 K{\o}benhavn {\O},
                 Denmark}

\altaffiltext{10}{Centre for Astrophysics and Cosmology,
                  Science Institute,
                  University of Iceland,
                  Dunhagi 5,
                  IS--107, Iceland}

\altaffiltext{11}{Department of Physics and Astronomy,
                  Uppsala University,
                  P. O. Box 516,
                  SE--751 20 Uppsala,
                  Sweden}

\altaffiltext{12}{NASA Ames Research Centre,
                  Moffett Field, CA 94035,
                  USA}

\altaffiltext{13}{Department of Physics and Astronomy,
                  University of Nevada,
                  Las Vegas, NV 89154,
                  USA}


\begin{abstract}

  GRB~090417B was an unusually long burst with a $T_{90}$ duration of
  at least 2130~s and a multi-peaked light curve at energies of
  15--150~keV.  It was optically dark and has been associated with a
  bright star-forming galaxy at a redshift of 0.345 that is broadly
  similar to the Milky Way.  This is one of the few cases where a host
  galaxy has been clearly identified for a dark gamma-ray burst and
  thus an ideal candidate for studying the origin of dark bursts.  We
  find that the dark nature of GRB~090417B cannot be explained by high
  redshift, incomplete observations, or unusual physics in the
  production of the afterglow.  Assuming the standard relativistic
  fireball model for the afterglow we find that the optical flux is at
  least 2.5~mag fainter than predicted by the $X$-ray flux.  The {\sl
    Swift\/}/XRT $X$-ray data are consistent with the afterglow being
  obscured by a dense, localized sheet of dust approximately 30--80~pc
  from the burst along the line of sight.  Our results suggest that
  this dust sheet imparts an extinction of $A_V \ga 12$~mag, which is
  sufficient to explain the missing optical flux.  GRB~090417B is an
  example of a gamma-ray burst that is dark due to the localized dust
  structure in its host galaxy.

\end{abstract}




\section{Introduction\label{SECTION:intro}}

The {\sl Swift\/} observatory \citep{GCC2004} is a multi-instrument
satellite mission that was designed to detect and rapidly localize
gamma-ray bursts (GRBs).  The observatory contains three telescopes:
the Burst Alert Telescope (BAT\@; \citet{BBC2005}), the $X$-Ray
Telescope (XRT\@; \citet{BHN2005}), and the UltraViolet/Optical
Telescope (UVOT\@; \citet{RKM2005}).  The BAT is used to identify GRBs
and localize them to $\sim 3\arcmin$ in the energy range 15--150 keV.
Once the BAT has localized a burst {\sl Swift\/} slews to point the
XRT and the UVOT at the burst.  The XRT obtains rapid $X$-ray
localizations to $\lesssim 5\arcsec$ in the energy range 0.2--10 keV
while the UVOT obtains localizations to $\approx 0\farcs5$ then
repeatedly cycles through a set of optical and ultraviolet filters.
Over the past five years {\sl Swift\/} has detected GRBs at the rate
of approximately 100 per year.  Almost all of these {\sl
  Swift\/}-detected GRBs have had $X$-ray afterglows, and optical or
infrared afterglows have been detected for $\approx 60$\% of them.
The remaining $\approx$ 40\% of the {\sl Swift\/} bursts have no
reported optical or infrared detections.  In general between $\approx
25$--40\% of GRBs are dark GRBs \citep{FJP2009}.

     The gamma-ray burst GRB~090417B was detected by the BAT as an
image trigger at 13:17:23 UT on 2009 Apr 17 \citep{SBB2009}.
{\sl Swift\/} immediately slewed to GRB~090417B and narrow-field
observations began observing 387~s (XRT) and 378~s (UVOT) after the
BAT trigger.  The BAT light curve showed a long period of emission
starting 200~s before the trigger (see Figure~\ref{FIGURE:batlc}).

Several ground-based follow-up observations were made, but no optical
or infrared afterglow was detected.  \citet{FMH2009} found a source
with $R = 21.3$ inside the XRT error circle \citep{SBB2009} and
identified it as an SDSS galaxy \citep{SWL2002} and the possible host
galaxy of GRB~090417B.  \citet{BF2009} measured a redshift of $z =
0.345$ for this galaxy based on the forbidden oxygen emission lines
[\ion{O}{2}](3727) and [\ion{O}{3}](5006) as well as H$\alpha$
emission.

Dark GRBs have been an enduring mystery in the lore of GRB studies.
In the pre-{\sl Swift\/} era only about one quarter of localized
gamma-ray bursts had an optical afterglow (OA) detected
\citep[\eg,][]{FJG2001}.  It was sometimes assumed that this was due
to the response times of ground-based optical telescopes which allowed
most afterglows to fade before they could be observed.  {\sl Swift\/}
however, with its ability to train the XRT and UVOT on a burst within
$\approx 100$~s of the initial detection, has been able to provide
positions to within $\approx 5\arcsec$ for most bursts that have been
detected by the BAT\@.  The rapid distribution of these positions has
allowed ground-based optical and infrared follow-up observations for
many of these bursts, often within minutes of the BAT trigger.  Even
with such rapid follow-up observations no optical or infrared
afterglows have been found for $\approx$ 40\% of BAT-detected GRBs,
thus demonstrating that observational constraints were not the cause
to the dark burst problem \citep{RSF2006}.

Several attempts have been made to quantify the dark burst problem by
establishing operational definitions of dark bursts
\citep{JHF2004,RWK2005,HKG2009}.  Most recently \citet{HKG2009}
proposed that dark GRBs be defined as those that have $\beta_{OX} <
\beta_X - 0.5$ measured 11 hours after the burst.  They define
$\beta_{OX}$ as the spectral index between the optical and $X$-ray
regimes and $\beta_X$ as the spectral index in the $X$-ray band.  The
spectral index is defined by $f_\nu(\nu) \propto \nu^{-\beta}$ where
$f_\nu(\nu)$ is the flux density at a frequency $\nu$.  This
definition assumes that both the $X$-ray and optical afterglows are
due to synchrotron radiation from a relativistic fireball.  The
synchrotron model predicts that the spectral slope will either be the
same in both regimes (if there is no cooling break between them) or
will differ by $\Delta\beta = 0.5$ (if there is a cooling break
between them).  A difference of $\Delta\beta > 0.5$ can not be
explained in this model and thus requires that something be
suppressing the flux at optical wavelengths, resulting in a dark burst
\citep{JHF2004}.

     Several possibilities have been proposed to explain dark bursts.
First, several studies have suggested that dark bursts may be due to
extinction along the line of sight, either in the Milky Way or in the
host galaxy \citep[\eg,][]{PFG2002,LFR2006,BFK2007,JRW2008,TLR2008}.
\citet{PCB2009} suggest that this is the case for most dark GRBs.
Further, $X$-ray observations show that some bursts with no optical
afterglows have systematically higher \ion{H}{1} column densities than
those with optical afterglows \citep{FJP2009}.  This suggests that
dark bursts may be suffering from higher extinction than
optically-bright bursts.  However, there is significant overlap in the
distribution of $N_{\mathrm{H}}$ values between dark and bright
bursts, so it is not clear if extinction is the sole parameter
responsible for the darkness of these bursts.

\citet{SMP2007,SPO2009} did panchromatic studies of several GRBs that
were observed with {\sl Swift\/}'s XRT and UVOT ({\ie},
optically-bright GRBs) in order to probe the extinction in their host
galaxies.  They found that the combined $X$-ray/ultraviolet/optical
data for most of these GRBs are best fit using a Small Magellanic
Cloud (SMC) dust model \citep{P1992}.  The extinctions in these
galaxies were small, with $A_V < 1$ mag.  They also found a mean
gas-to-dust ratio of $N_\mathrm{H} = \left(6.7 \times 10^{21}\right)
A_V$, which is lower than, but similar to, the value found in the SMC
($N_\mathrm{H} = \left(15.4 \times 10^{21}\right) A_V$ using equation
(4) and Table 2 of \citet{P1992}).  \citet{KKZ2006} also found that
the the gas-to-dust ratio for a large sample of host galaxies of
optically-bright GRBs is lower than the SMC value.  However, the
gas-to-dust ratios in the host galaxies of dark GRBs tend to be more
like that of the Milky Way than that of the SMC
\citep[\eg,][]{PFG2002,BFK2007,JRW2008,EFH2009}.

A second possible origin for dark bursts is that they are GRBs located
at high redshift, so the Lyman break (or the start of the Lyman-alpha
forest) is observed redward of the optical band.  The most distant GRB
observed to date is GRB~090423 at $z = 8.2$ \citep{SDC2009,TFL2009},
so some of the dark bursts could simply be at high redshift.  However,
\citet{FJP2009} have found that less than approximately 19\% of GRBs
have $z > 7$, whereas between 25\% and 42\% of GRBs are dark.
Therefore, it is unlikely that all of the {\sl Swift\/} dark burst lie
at very high redshifts.  The affects of relativistic beaming of GRB
emission at very high redshift is poorly understood due to a lack of
data on high-redsift bursts.  We assume here that it does not affect
our discussion in this paper.

     Third, the hypothesis that afterglow radiation is synchrotron
radiation may be wrong.  This could lead to a different spectrum from
what is expected from the relativistic fireball model, in which case the
above definition of a dark burst is invalid.

     Long--soft GRBs \citep{KMF1993} tend to be found in host galaxies
that are small, irregular, and have high specific star-formation rates
\citep{CHG2004}.  \citet{LDM2003} found a median infrared luminosity of
$L \approx 0.08 L^*$ for the host galaxies of GRBs with optical
afterglows and conclude that most hosts are sub-luminous at optical
and near-infrared wavelengths.  These GRB hosts tend to be small
galaxies that have intrinsic luminosities similar to those of the dwarf
galaxies in the Local Group.  However, some hosts are significantly
larger, such as that of GRB~990705 \citep{LDM2002}, which was an Sc
spiral with $L \approx 2L_\star$.  To date no long--soft GRB has been
found in an elliptical galaxy.  All of the hosts of long--soft GRBs
have morphologies that are consistent with either exponential discs,
or irregular structure \citep{CVF2005,WBP2007}.  GRB hosts follow the
size--luminosity relation \citep[\eg,][]{TFR2006} and extend it
towards lower luminosities \citep{WBP2007}.  In spite of their
generally small size the host galaxies of long--soft GRBs tend to have
high specific star-formation rates (SSFRs).  \citet{SGL2009} found a
median SSFR of 0.8 yr$^{-1}$, similar to what is seem in Lyman-break
galaxies.

It is important to realize, however, that these properties have been
determined using GRBs that have usually been localized to
sub-arcsecond precisions.  In practice this usually means that the
sample of GRB host galaxies is restricted to GRBs that have had
optical afterglows.  The host galaxies of dark GRBs may have different
properties.  \citet{PCB2009} identify host galaxies for 14 dark GRBs
and found that they have the same redshift distribution as the hosts of
optically-bright GRBs.  They also found that these hosts do not appear to be
significantly different in size, structure, or luminosity from the
hosts of GRBs with optical afterglows.  They did find that a
significant fraction of the hosts of dark bursts show evidence for
high internal extinction ($A_V > 2$--5 mag).  However, many of their
host associations are uncertain, with more than half of the galaxies
having a probability of a chance association on the sky of
$P_\mathrm{ch} > 0.01$.  As we will show in this paper GRB~090417B is
located $1\farcs08$ from the centre of a bright galaxy with a
probability of a chance association of only $P_\mathrm{ch} \approx
10^{-3}$.  This makes GRB~090417B one of the strongest associations of
a dark GRB with a host galaxy.

In this paper we present space- and ground-based gamma-ray, $X$-ray,
ultraviolet, and optical observations of GRB~090417B and its host
galaxy.  We have adopted a standard ${\Lambda}$CDM cosmology with a
Hubble parameter of $H_0 = 71$ km s$^{-1}$ Mpc$^{-1}$, a matter
density of $\Omega_m = 0.27$, and a cosmological constant of
$\Omega_\Lambda = 1 - \Omega_m = 0.73$.  For this cosmology a redshift
of $z = 0.345$ corresponds to a luminosity distance of 1.806 Gpc and a
distance modulus of 41.29 mag.  One arcsecond corresponds to 6.51
comoving kpc, or 4.84 proper kpc.  The look back time is 3.80 Gyr.


\section{Data\label{SECTION:data}}


\subsection{BAT Data\label{SECTION:bat_data}}

     The BAT scaled-map, event and DPH data were reduced using the
standard BAT software available through
HEASARC\footnote{http://swift.gsfc.nasa.gov/docs/software/lheasoft/download.html}.

     The $T_{90}$ duration of GRB~090417B was $>2130 \pm 50$~s in the
15--150~keV band. The light curve evolved gradually, and there were no
features in the unimaged count rates that might indicate how long the
burst persisted in the BAT energy range past that time.  Even at that
lower limit on $T_{90}$, this was the longest GRB ever seen by BAT or
BATSE \citep{FM1989}, although GRB~060218 approached this duration
with $T_{90} = 2100 \pm 100$ \citep{CMB2006}.

\begin{figure}
  \rotate
  \includegraphics[angle=+90,scale=0.7]{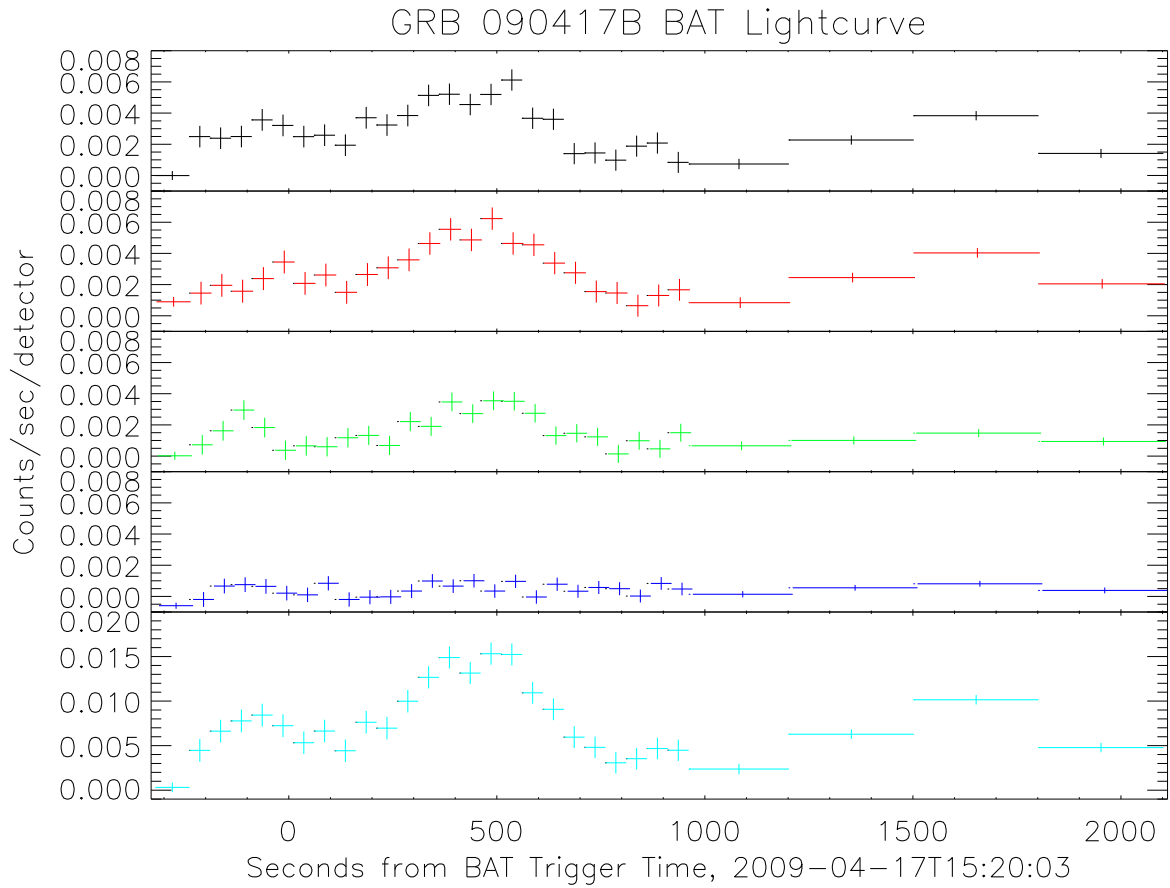}

  \figcaption{The top four panels show the BAT light curves for the
    prompt emission from GRB~090417B in four energy bands.  From the
    top down the energy bands are: 14--24~keV, 24--51.1~keV,
    51.1--101.2~kev, and 101.2--194.9~keV.  The bottom panel shows the
    total light curve in the 14--194.9~keV band.\label{FIGURE:batlc}}
\end{figure}

     The BAT light curve for GRB~090417B showed four broad,
overlapping peaks in the flux at approximately $T - 70$, $T + 400$, $T
+ 500$, and $T + 1600$~s, as shown in Figure~\ref{FIGURE:batlc}. The
1-s peak flux of $0.3 \pm 0.1$ photons cm$^{-2}$ s$^{-1}$ in the range
15--350 keV was approximately $T + 500$~s.  The total fluence from $T
- 320$ to $T + 2105$~s was $8.20^{+1.0}_{-2.1} \times 10^{-6}$ erg
cm$^{-2}$ in the 15--150~keV band.  The spectrum was best fitted by a
simple power-law function with an average photon index of $1.89 \pm
0.12$.  The index did not change significantly during the burst.  All
the above uncertainties are at 90\% confidence.

     The most impressive characteristic in the prompt emission of
GRB~090417B was its extreme length. The burst was marginally above
background ($2.3\sigma$) on the 64-s interval starting immediately
after a pre-planned slew maneuver ended at $T - 320$ seconds, although
not at a high enough level to be detectable independent of the later
emission.  From that time, the flux increased steadily, to be detected
as a point source on-board on the interval $T - 64$ to $T_0$~s, and
finally as a sufficiently significant source to trigger a burst
response on the interval $T_0$ to $T + 320$~s.  Although it appeared
to be declining toward background levels, the burst continued to be
detected until the spacecraft slewed away due to the changing Earth
constraint at $T + 2105$~s.


\subsection{XRT Data\label{SECTION:xrt_data}}


\subsubsection{Data Reduction\label{SECTION:xrt_data_reduction}}

     The XRT on board {\sl Swift\/} began observations on 2009 April
17 at 15:26:33 UT, 387~s after the BAT trigger \citep{SBB2009}, and
ended on 2009 May 5, with a total net exposure of 1456~s in Windowed
Timing mode (WT) and 83.36~ks in Photon Counting mode (PC). The XRT
observations are listed in Table~\ref{TABLE:xrtjournal}.

\begin{deluxetable}{lllllcr}
  \tabletypesize{\scriptsize}

  \tablecaption{Journal of XRT observations.\label{TABLE:xrtjournal}}
  
  \tablehead{%
    \colhead{Sequence} &
    \colhead{Start Time} &
    \colhead{End Time} &
    \colhead{Start time\tablenotemark{1}} &
    \colhead{End Time\tablenotemark{1}} &
    \colhead{Obs.\ Mode\tablenotemark{2}} &
    \colhead{Exposure} \\
    \colhead{} &
    \colhead{UT} &
    \colhead{UT} &
    \colhead{(s)} &
    \colhead{(s)} &
    \colhead{} &
    \colhead{(s)}
  }
  
  \startdata 
  00349450000 & 2009-04-17 15:26:40 & 2009-04-17 23:29:46  & $3.932 \times 10^{2}$ & $2.055 \times 10^{3}$ & WT  & $1.456 \times 10^{3}$ \\
  00349450000 & 2009-04-17 15:38:43 & 2009-04-17 23:33:26  & $1.116 \times 10^{3}$ & $2.960 \times 10^{4}$ & PC  & $9.214 \times 10^{3}$ \\
  00349450002 & 2009-04-18 10:29:48 & 2009-04-18 11:07:52  & $6.898 \times 10^{4}$ & $7.124 \times 10^{4}$ & PC  & $2.252 \times 10^{3}$ \\
  00349450003 & 2009-04-18 16:57:16 & 2009-04-18 20:20:44  & $9.223 \times 10^{4}$ & $1.044 \times 10^{5}$ & PC  & $4.871 \times 10^{3}$ \\
  00349450006 & 2009-04-19 01:17:02 & 2009-04-19 12:50:57  & $1.222 \times 10^{5}$ & $1.638 \times 10^{5}$ & PC  & $9.709 \times 10^{3}$ \\
  00349450007 & 2009-04-20 04:41:58 & 2009-04-20 22:26:57  & $2.209 \times 10^{5}$ & $2.848 \times 10^{5}$ & PC  & $6.131 \times 10^{3}$ \\
  00349450008 & 2009-04-21 14:13:55 & 2009-04-21 23:59:56  & $3.416 \times 10^{5}$ & $3.768 \times 10^{5}$ & PC  & $7.910 \times 10^{3}$ \\
  00349450009 & 2009-04-22 04:29:42 & 2009-04-22 14:20:47  & $3.930 \times 10^{5}$ & $4.284 \times 10^{5}$ & PC  & $4.574 \times 10^{3}$ \\
  00349450010 & 2009-04-23 00:07:52 & 2009-04-23 22:40:57  & $4.637 \times 10^{5}$ & $5.448 \times 10^{5}$ & PC  & $2.989 \times 10^{3}$ \\
  00349450011 & 2009-04-24 00:18:14 & 2009-04-24 22:57:56  & $5.507 \times 10^{5}$ & $6.323 \times 10^{5}$ & PC  & $1.344 \times 10^{3}$ \\
  00349450012 & 2009-04-25 11:39:16 & 2009-04-25 23:01:57  & $6.779 \times 10^{5}$ & $7.189 \times 10^{5}$ & PC  & $1.480 \times 10^{3}$ \\
  00349450013 & 2009-04-26 02:11:18 & 2009-04-27 23:12:56  & $7.303 \times 10^{5}$ & $8.923 \times 10^{5}$ & PC  & $2.140 \times 10^{3}$ \\
  00349450014 & 2009-04-27 00:40:17 & 2009-04-27 21:39:56  & $8.112 \times 10^{5}$ & $8.868 \times 10^{5}$ & PC  & $1.224 \times 10^{3}$ \\
  00349450015 & 2009-04-28 00:37:57 & 2009-04-28 23:20:56  & $8.975 \times 10^{5}$ & $9.792 \times 10^{5}$ & PC  & $8.368 \times 10^{3}$ \\
  00349450016 & 2009-04-29 00:41:42 & 2009-04-29 23:26:57  & $9.841 \times 10^{5}$ & $1.066 \times 10^{6}$ & PC  & $4.537 \times 10^{3}$ \\
  00349450017 & 2009-05-01 00:40:42 & 2009-05-02 23:37:57  & $1.157 \times 10^{6}$ & $1.326 \times 10^{6}$ & PC  & $1.604 \times 10^{4}$ \\
  \enddata
  
  \tablenotetext{l}{Time since BAT trigger.}
  \tablenotetext{2}{WT, window timing mode; PC, photon counting mode}
\end{deluxetable}

     The XRT data were processed using the FTOOLS software package
distributed inside HEASOFT (v6.6.3). We ran the task {\sc xrtpipeline}
applying calibrations and standard filtering criteria.  Events with
grades 0--2 and 0--12 were selected for WT and PC data
respectively. The analysis was performed inside the 0.3--10~keV energy
band.

     The best position for the afterglow was obtained using 6972~s of
overlapping XRT Photon Counting mode data and 9 UVOT $v$-band images
in order to correct the XRT astrometry making use of the XRT--UVOT
alignment and matching to the USNO-B1 catalogue, as described by
\citet{GTB2007} and \citet{EBP2009}.  Our best position is RA(J2000.0)
= $209\fdg6942$ (13:58:46.62), Dec(J2000.0) = $+47\fdg0182$
(+47:01:05.4) with an error of $1\farcs4$ (90\% confidence, including
boresight uncertainties).


\subsubsection{Temporal Analysis\label{SECTION:xa_temporal}}

\begin{figure}
  \rotate
  \includegraphics[angle=-90,scale=0.7]{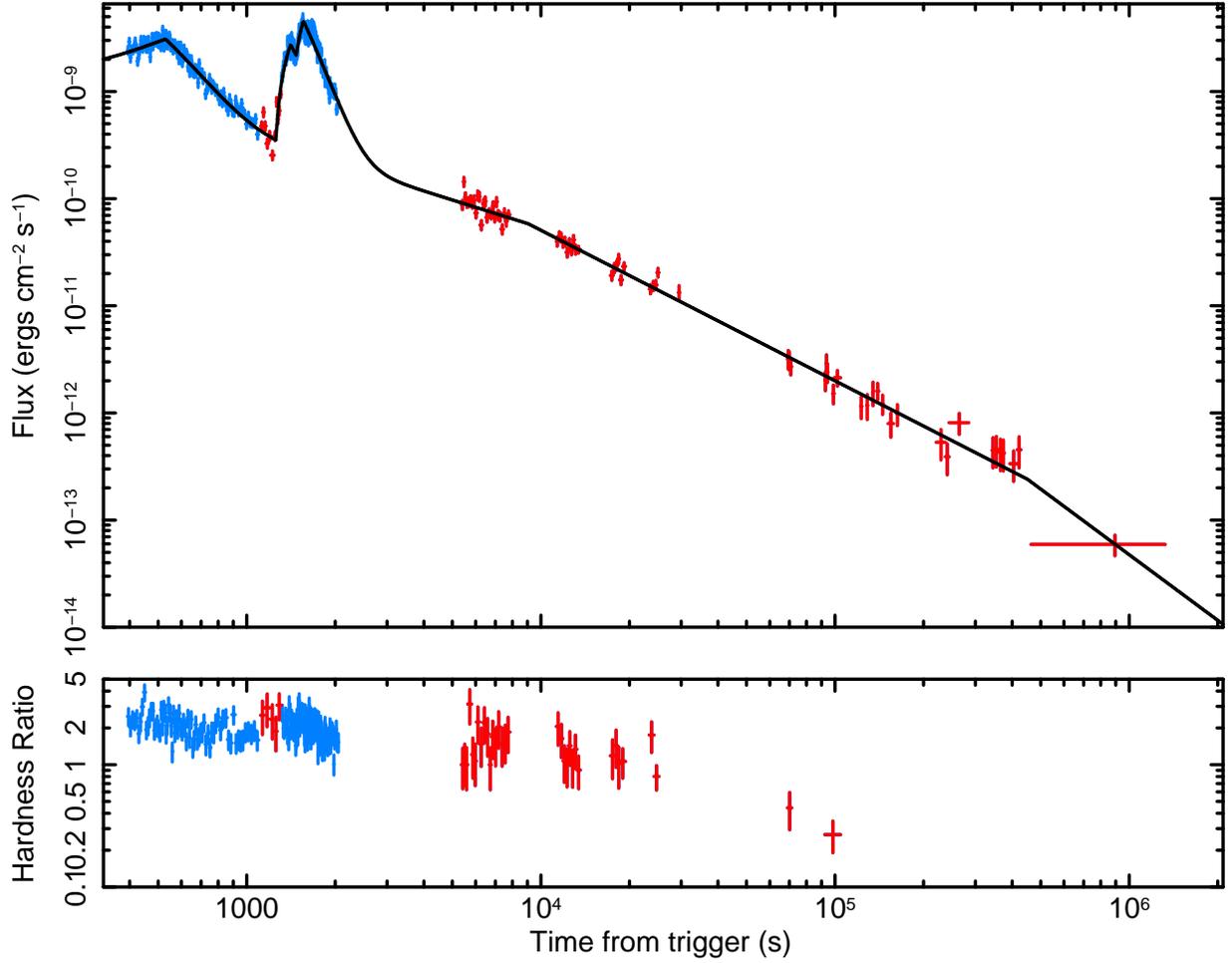}

  \figcaption{Top: The $X$-ray flux light curve of
    GRB~090417B. Bottom: The hardness ratio evolution of GRB~090417B
    in the 0.3--2 and 2--10~keV bands.  The blue points are window
    timing mode data and the red points are photon counting mode
    data.\label{FIGURE:xa_lcurve}}
\end{figure}

Source photons were extracted from a region with a 30 pixel radius (1
pixel = $2\farcs36$), with the exception of PC mode data for sequence
00349450000 where we used an annular region with radii 3 pixel and 30
pixel in order to correct for pile-up. The background was estimated
from a circular region with a 50 pixel radius located away from any
detected source in the field. When the count-rate dropped below the
level of $\approx 10^{-2}$ count s$^{-1}$ we used the SOSTA tool of
{\sc ximage}, which corrects for vignetting, exposure variations and
PSF losses within an optimized box, using the same background region.
The 0.3--10~keV light curve thus obtained is shown in
Figure~\ref{FIGURE:xa_lcurve} (top panel). The light curve has been
rebinned in order to achieve a minimum signal-to-noise ratio (S/N) of
3 for each point. The best fit to the light curve is given by a doubly
broken power law with indices $\alpha_1 = 0.86 \pm 0.03$, $\alpha_2 =
1.40 \pm 0.04$, and $\alpha_3 = 2.0^{+0.7}_{-0.4}$, and breaks at
$T+9100^{+900}_{-1000}$~s and $T+4.5^{+2.8}_{-1.2} \times 10^{5}$~s.
The second break could be interpreted as a jet break.  However the
lack of an optical detection for this burst does not allow us to
confirm or deny this hypothesis.  Also see
\S~\ref{SECTION:dust_scattering} for an alternate interpretation of
the late-time steepening of the $X$-ray light curve.

The first sequence of $X$-ray data is dominated by a strong flare.
There is a rise that began before the start of XRT observations and
peaked around $T+530$~s. This is followed by a double-peaked flare
which started at $T+1260$~s, peaked at $T+1410$~s and $T+1470$~s, and
had a decay time of 240~s.  Further flaring activity is apparent up to
at least $T+4 \times 10^{5}$~s.  The bottom panel of
Figure~\ref{FIGURE:xa_lcurve} shows the hardness ratio for the
afterglow measured using the 0.3--2 and the 2--10~keV bands.  The
first orbit (up to $T+1300$) shows a hard hardness ratio with spectral
evolution that follows the flaring activity.  The same is true for the
remaining observations, with the hardness ratio slowly decreasing with
time.

\subsubsection{Spectral Analysis\label{SECTION:xa_spectral}}

     In order to detect possible spectral variations we extracted
spectra separately for the two flares, the early decay part from
$T+5000$~s to $T+9100$~s, the steep decay from $T+9100$~s to $T+4.5
\times 10^5$~s, and the late decay from $T+4.5 \times 10^5$~s to the
end of observations.  Source and background spectra were extracted
using the same regions used for the light curves, with the exception
of the late decay spectrum where the count rate was too low to group
the energy channels in order to have a Gaussian distribution of the
number of photons per grouped channel.  Hence, for this spectrum we
replaced the $\chi^2$ statistic with the $C$ statistic \citep{C1979},
which can be used whenever the Gaussian approximation is not valid,
provided that the contamination from background photons is negligible.
To ensure this, we extracted photons from a region with a 10 pixel
radius.

The ancillary response files were generated using the task {\sc
  xrtmkarf}, and with the exception of the late decay spectrum we
grouped channels in order to have at least 20 photons per bin.
Spectral fitting was performed using {\sc XSpec} (v11.3.2).  For each
spectrum we modelled the Galactic absorption using a warm absorber
with neutral hydrogen column density $1.60 \times 10^{21}$ cm$^{-2}$
\citep{KBH2005} and included an intrinsic warm absorber at the
redshift of the host galaxy ($z =0.345$).  The best model for the
spectra of the flares was a power law with a high energy cutoff, while
for the remaining parts of the afterglow the best fit was obtained
using a single power law. The best fit parameters are given in
Table~\ref{TABLE:xa_spectra}.

\begin{deluxetable}{lllllll}
  \tabletypesize{\scriptsize}

  \tablecaption{Spectral parameters.\label{TABLE:xa_spectra}}

  \tablehead{%
    \colhead{Start time\tablenotemark{1}} &
    \colhead{End Time\tablenotemark{1}} &
    \colhead{$N_{\mathrm{H}}$} &
    \colhead{$\Gamma$} &
    \colhead{Cutoff Energy} &
    \colhead{Mean Flux} &
    \colhead{$\chi^2/$d.o.f.} \\
    \colhead{(s)} &
    \colhead{(s)} &
    \colhead{($10^{22}$ cm$^{-2}$)} &
    \colhead{} &
    \colhead{(keV)} &
    \colhead{(erg cm$^{-2}$ s$^{-1}$)} &
    \colhead{}}
  
  \startdata
  $3.932 \times 10^{2}$ & $1.116 \times 10^{3}$ & $1.5 \pm 0.2$ & $0.2 \pm 0.2$ & $3.0^{+0.7}_{-0.5}$ & $1.3 \times 10^{-09}$ & 446/421 \\
  $1.116 \times 10^{3}$ & $2.055 \times 10^{3}$ & $1.3 \pm 0.2$ & $0.3 \pm 0.2$ & $2.9^{+0.5}_{-0.4}$ & $2.8 \times 10^{-09}$ & 604/523 \\
  $5.000 \times 10^{3}$ & $9.100 \times 10^{3}$ & $2.0 \pm 0.4$ & $1.9 \pm 0.2$ & \nodata           & $6.2 \times 10^{-11}$ &  39/45 \\
  $9.100 \times 10^{3}$ & $2.960 \times 10^{}$ & $1.6 \pm 0.4$ & $2.3 \pm 0.2$ & \nodata           & $1.7 \times 10^{-11}$ &  15/29 \\
  $6.898 \times 10^{4}$ & $4.284 \times 10^{5}$ & $1.9 \pm 0.5$ & $3.8 \pm 0.5$ & \nodata           & $4.4 \times 10^{-13}$ &  13/22 \\
  $4.637 \times 10^{5}$ & $1.326 \times 10^{6}$ & $<1.9$        & $3.1 \pm 0.9$ & \nodata           & $7.7 \times 10^{-14}$ &  32/24\tablenotemark{2} \\
  \enddata

  \tablenotetext{1}{Time since BAT trigger.}
  \tablenotetext{2}{$C$-statistic/bins}
\end{deluxetable}

     The fact that the first phases of the XRT observations are best
fitted by a cutoff power law with the cutoff energy around 3~keV
indicates that the peak energy of the flares was passing through the
XRT band at the time.  After the flares the spectrum is described by a
power law with photon index of $2.0 \pm 0.1$.  After the first break,
up to $T+29.6$~ks, the spectrum is described by a power law with
similar parameters ($\Gamma = 2.3 \pm 0.2$ and an absorbing column
$N_{\mathrm{H}} = 2.0 \pm 0.4 \time 10^{22}$ cm$^{-2}$). At later
times the spectrum softens significantly to a photon index of $\Gamma
= 3.8 \pm 0.5$ and shows no significant variations across the break at
$T + 4.5 \times 10^5$~s up to the end of the observations.

     The intrinsic absorbing column is in the range $(1.1-2.4) \times
10^{22}$ cm$^{-2}$.  From our analysis of the spectral energy
distribution (\S~\ref{SECTION:sed}) we find that the best fit to the
data is given by a Milky Way extinction law instead of the SMC-like
extinction law found in many GRB host galaxies.  Therefore, we use the
relationship between $N_\mathrm{H}$ and extinction in the Milky Way
from \citet{PS1995} to find that the observed absorbing column
corresponds to an extinction $A_v \approx 11$~mag.  This high value
for the extinction in the host along the line of sight to the burst
is capable of explaining the dark optical nature of this event.


\subsection{UVOT Data\label{SECTION:uvot_data}}

     The {\sl Swift\/}/UVOT began observations 378~s after the BAT
trigger \citep{SBB2009} with a 9~s settling mode exposure using the
$v$ filter.  This was followed by a sequence that rotated through all
seven of UVOT's lenticular filters.  The first of these exposures was
a 147~s exposure starting at 395~s taken with the white filter.
Observations continued until 1\,325\,874~s (= 15.3~days) after the
trigger using primarily the $u$ and ultraviolet filters.  No optical
or ultraviolet afterglow was detected in any of the UVOT data.  Once
it was determined that there was no optical or ultraviolet afterglow
for GRB~090417B UVOT observed primarily with its ultraviolet filters
in order to obtain ultraviolet magnitudes for the nearby galaxy that
can not be obtained from ground-based observatories.

\begin{figure}

\includegraphics[scale=0.95,angle=-90]{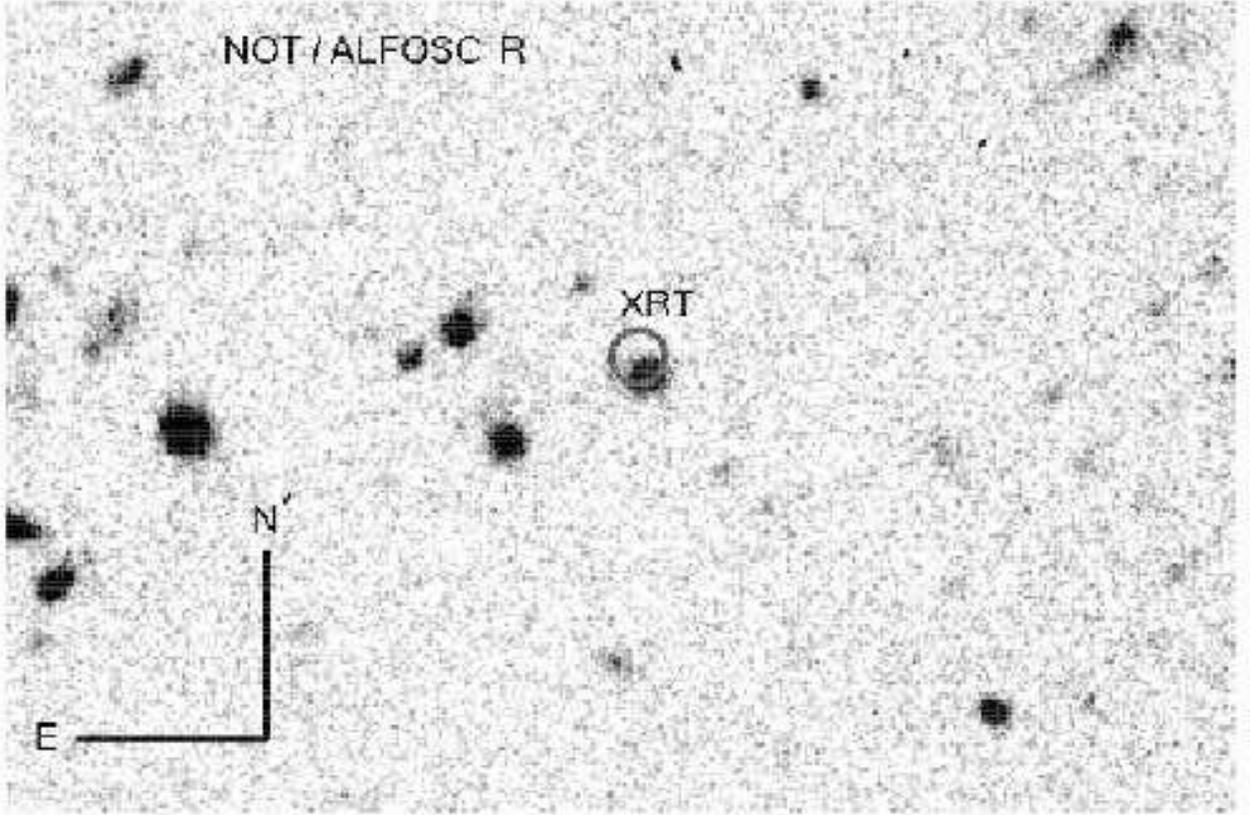}

\figcaption[./uvot_fc2.ps]{This Figure shows the averaged NOT/ALFOSC
  $R$-band image of the field of GRB~090417B with the XRT error
  circle.  The SDSS galaxy is in the southeast quadrant of the XRT
  error circle.  The arms of the compass have a length of ten
  arcseconds.\label{FIGURE:not_field}}
\end{figure}

The SDSS galaxy noted by \citet{FMH2009} is detected when we coadd the
UVOT images.  We determined a centroid for the SDSS galaxy of RA, Dec
= 13:58:46.66, +47:01:04.4 (J2000.0) with an estimated internal
uncertainty of $0\farcs67$ and an estimated systematic uncertainty
relative to the USNO-B1.0 catalogue of $0\farcs42$ \citep{BCH2010}.
These uncertainties are the 90\% confidence intervals.  This galaxy is
$1\farcs08$ southeast of the centre of the UVOT-enhanced XRT error
circle.  The field of GRB~090417B is shown in
Figure~\ref{FIGURE:not_field}.  The SDSS galaxy is well-isolated from
other sources in the field, so there is no contamination from
neighbouring sources when doing aperture photometry.

We obtained the UVOT data from the {\sl Swift\/} Data Archive on 12
May 2009.  This data has had bad pixels identified, mod-8 noise
corrected, and has been transformed into FK5 coordinates.  We used the
standard UVOT data analysis software distributed with HEASOFT 6.6.2
along with the standard calibration data.  Photometry was done on the
SDSS galaxy using {\sc uvotsource} with a circular source aperture of
radius $4\farcs0$ for the source region and a circular aperture of
radius $15\arcsec$ centred at RA, Dec = 13:58:47.57, +47:00:33.3
(J2000.0) for a background region.  The background region was selected
to have similar background properties to those at the location of the
galaxy, and to be free of contaminating sources.  The UVOT photometry
of the SDSS galaxy is presented in Table~\ref{TABLE:uvot_photometry}.
We find no evidence for any change in the magnitude of the galaxy, in
any filter, during the course of the UVOT observations.  Our
photometry was calibrated to the UVOT photometric system described in
\citet{PBP2008}.

\begin{deluxetable}{crcc}

\tablecaption{This Table lists the UVOT photometry of
  SDSS~J135846.65+470104.5, the host galaxy of GRB~090417B.  Exposure
  is the total exposure time in each filter.  All upper limits are
  3-$\sigma$ upper limits.\label{TABLE:uvot_photometry}}

\tablehead{%
        \colhead{Filter} &
        \colhead{Exposure} &
        \colhead{Mag} &
        \colhead{Err}
}

\startdata
$v$   &    1237 & $>$20.9  & \nodata \\
$b$   &     862 & $>$21.4  & \nodata \\
$u$   & 22\,554 &    23.09 & 0.38    \\
uvw1  & 10\,850 &    23.12 & 0.56    \\
uvm2  & 15\,942 &    22.49 & 0.29    \\
uvw2  & 30\,827 &    22.69 & 0.19    \\
white &     841 & $>$22.2  & \nodata \\
\enddata

\end{deluxetable}

The UVOT-enhanced XRT position for GRB~090417B is RA = 13:58:46.62,
Dec = +47:01:05.4 (J2000.0) which corresponds to Galactic coordinates
of $\ell^\mathrm{II},b^\mathrm{II} = 93\fdg7486,+66\fdg1127$.  The
line-of-sight Galactic reddening in this direction is $E_{B\!-\!V} =
0.02 \pm 0.01$ mag \citep{SFD1998}.  This corresponds to extinctions
in the UVOT filters of of $A_v = 0.05$, $A_b = 0.07$, $A_u = 0.08$,
$A_{uvw1} = 0.11$, $A_{uvm2} = 0.16$, and $A_{uvw2} = 0.14$, and
$A_{white} = 0.08$ mag.

Since there is a candidate host galaxy for GRB~090417B it is not
possible to put a direct faint limit on the luminosity of the optical
afterglow beyond the observed magnitude of the galaxy itself ($R =
21.3$ \citet{FMH2009}).  In order to constrain the maximum possible
luminosity of the afterglow we used the UVOT data to look for
variability in the galaxy's magnitude.  To do this we divided the data
into early and late epochs.  The early epoch consisted of all the UVOT
data taken during {\sl Swift\/}'s initial automated observations of
GRB~090417B (OBSID 00349450000), which covers times up to 29\,604~s
($\approx$ 8~hr) after the BAT trigger.  This is the period when the
optical afterglow is expected to be brightest.  The late epoch
included all observations later than this.  The UVOT exposures for
each epoch were coadded to produce two deep images for each UVOT
filter.  We photometered the galaxy in each image and recorded the
difference between the two magnitudes.  If we assume that the first
epoch magnitude is due to the host+afterglow and the second epoch
magnitude is due to the host alone then the difference between these
two magnitudes puts an upper limits on the luminosity of the
afterglow.  We find no evidence for any change in the magnitude of
the host galaxy between the first and last epochs larger than $\pm
0.2$ mag.  Therefore, the afterglow could not have contributed more
light than what corresponds to a 0.2 mag increase in luminsity.  From
this we obtain the following constraints on the magnitude of the
optical afterglow at 4~hr (= 14\,800~s) after the BAT trigger:
$u_\mathrm{oa} > 24.9$, uvw1$_\mathrm{oa} > 24.9$, uvm2$_\mathrm{oa} >
24.1$, and uvw2$_\mathrm{oa} > 24.4$.  The uncertainty in these
estimates is about 1.0 mag.


\subsection{Nordic Optical Telescope Data\label{SECTION:not_data}}

We obtained optical images of GRB~090417B on 2009 Apr 17--18 with the
ALFOSC-FASU mounted on the Nordic Optical Telescope (NOT).
ALFOSC-FASU was operated in its high gain state (0.726 e$^-$/ADU).
The read-out noise was 3.2 e$^-$/pixel and the pixel scale was
$0\farcs19$/pixel.  The NOT $R$-band image is shown in
Figure~\ref{FIGURE:not_field}.  Data reduction was done following the
standard procedure for optical CCD data and the individual images were
aligned by centroiding on several stars in each exposure.  Photometry
was performed on the combined NOT images using {\sc SExtractor}
\citep{BA1996}.  We calibrated the data using the standard ALFOSC zero
points\footnote{http://www.not.iac.es/instruments/alfosc/zpmon/}.  We
find $B = 23.25 \pm 0.13$ (Bessel $B$), $R = 21.34 \pm 0.03$ (Bessel
$R$), and $i = 20.82 \pm 0.07$ (interference $i$) for the SDSS galaxy.
The Galactic extinctions for the NOT data are $A_B = 0.07$, $A_R =
0.05$, \citep{SFD1998}, and $A_i =
0.03$\footnote{http://irsa.ipac.caltech.edu/applications/DUST/}.


\section{The Host Galaxy\label{SECTION:host}}


\subsection{Probability that the SDSS Galaxy is the Host\label{SECTION:alignment}}

     We used the methodology of \citet{BKD2002} to estimate the
probability that a galaxy would lie within $1\farcs08$ of the centre
of the XRT error circle by chance, $P_\mathrm{ch}$.  This assumes that
galaxies are randomly distributed on the sky and that there is no
clustering.  This assumption is not formally correct as galaxies do
cluster.  However, an examination of the field of GRB~090417B suggests
that this GRB did not occur in a significant over-density of
background galaxies, so distortions in the computed $P_\mathrm{ch}$
value are likely to be small.  \citet{BKD2002} use the galaxy number
counts of \citet{HPM1997} to generate the surface density of galaxies
with $R$ magnitudes brighter than some limiting magnitude.  In this
case the $R$ band is for the Keck 10-m Low Resolution Imaging
Spectrometer (LRIS) $R$ filter.  We adopt $R = 21.34 \pm 0.03$
(\S~\ref{SECTION:not_data}) as the magnitude of the SDSS galaxy.
Using the \citet{BKD2002} formula for an error circle containing a
galaxy this yields $P_\mathrm{ch} = 0.003$.  This is a fairly small
value, which suggests that SDSS~J135846.65+470104.5 is likely to be
the host galaxy of GRB~090417B.

A chance alignment probability of $P_\mathrm{ch} \approx 10^{-3}$ is
large enough that one can not rule out the possibility of a
misidentification.  A visual examination of the NOT images shows that
the XRT error circle includes a significant fraction of the SDSS
galaxy (see Figure~\ref{FIGURE:not_field}).  The observed separation
between the $X$-ray afterglow and the centre of the galaxy is
$1\farcs08$.  At a redshift of $z = 0.345$ this corresponds to a
projected separation of 5.23 proper kpc.  The galaxy appears
point-like in all of our data.  The NOT $R$-band data has a seeing
FWHM of $0\farcs91$, which represents an upper limit on the effective
radius of the galaxy of $r_e \la 4.4$ proper kpc \citet{BFT2009} found
$g^\prime = 22.47 \pm 0.16$ for the galaxy, which implies an absolute
magnitude of $M_V \approx -18.8$.  This puts the galaxy inside the
observed range of GRB host sizes and luminosities (see
\citet{FLS2006}, their Fig.~4).

     In order to test our hypothesis that SDSS~J135846.65+470104.5 is
the host galaxy of GRB~090417B we obtained the {\sl Chandra\/}
observations of the $X$-ray afterglow that were taken on 2009 May 11,
24 days ($\approx 2 \times 10^6$~s) after the burst (Proposal Number:
10900117, PI\@: Andrew Levan).  An $X$-ray source (consistent with the
XRT position) is clearly detected at the position of the SDSS galaxy.
The observed flux from this source is consistent with the $X$-ray flux
that is predicted from the XRT light curve assuming a power-law decay
with $\alpha_3 = 2$.  This suggests that the {\sl Chandra\/} source is
the $X$-ray afterglow of GRB~090417B, and that there is minimal (if
any) contamination from the galaxy.

     Taken together these arguments strengthen the case that
SDSS~J135846.65+470104.5 is the host galaxy.  Therefore, we conclude
that we have identified the host of the dark burst GRB~090417B to a
high degree of confidence.


\subsection{Spectral Energy Distribution\label{SECTION:sed}}

     We used the combined ultraviolet, optical, and infrared
photometry of the SDSS galaxy to determine its spectral energy
distribution (SED).  Photometry for this galaxy was taken from our
observations, the SDSS Web site\footnote{http://www.sdss.org/DR7/},
and the various GCN Circulars on this burst, and is listed in
Table~\ref{TABLE:host_photometry}.  We used {\sc XSpec} to fit a power
law to the observed SED\@.  The best fit, with no host extinction, has
a power-law index of $\Gamma = 2.67^{+0.13}_{-0.03}$.  However, the
goodness of fit is only $\chi^2/\mathrm{dof} = 81.40/12$, which is
very poor.  Therefore, we also tried including extinction due to dust
in the host galaxy when fitting the SED\@.  Dust laws for the Small
Magellanic Cloud, Large Magellanic Cloud and the Milky Way were tried.
These led to a significant improvement in the goodness of fit with the
best fit being found for Milky Way extinction in the host galaxy.  Our
best fit has a spectral index of $\Gamma \approx
0.80^{+1.19}_{-0.55}$, and a host extinction of $A_V =
3.5^{+1.0}_{-0.5}$.

\begin{deluxetable}{cccl}

\tablecaption{This Table lists all of the available photometry of
  SDSS~J135846.65+470104.5, the host galaxy of
  GRB~090417B.\label{TABLE:host_photometry}}

\tablehead{%
        \colhead{Filter} &
        \colhead{Mag} &
        \colhead{Err} &
        \colhead{Source}
}

\startdata
uvw2             &    22.69 &  0.19   &  1 \\
uvm2             &    22.49 &  0.29   &  1 \\
uvw1             &    23.12 &  0.56   &  1 \\
$u_\mathrm{UVOT}$ &    23.09 &  0.38   &  1 \\
$u^\prime$        &    23.23 &  0.53   &  2 \\
$b_\mathrm{UVOT}$ & $>$21.4  & \nodata &  1 \\
$B$              &    23.25 &  0.13   &  1 \\
$g^\prime$        &    22.02 &  0.16   &  2 \\
$g^\prime$        &    22.47 &  0.16   &  3 \\
$v_\mathrm{UVOT}$ & $>$20.9  & \nodata &  1 \\
$R$              &    21.34 &  0.03   &  1 \\
$r^\prime$        &    21.9  &  0.3    &  4 \\
$r^\prime$        &    21.62 &  0.09   &  2 \\
$r^\prime$        &    21.62 &  0.10   &  3 \\
$i$              &    20.82 &  0.07   &  1 \\
$i^\prime$        &    21.41 &  0.11   &  2 \\
$i^\prime$        &    21.31 &  0.12   &  3 \\
$z^\prime$        &    20.78 &  0.25   &  2 \\
$z^\prime$        &    21.49 &  0.30   &  3 \\
$J$              &    20.0  & \nodata &  5 \\
$K_s$            &    18.5  & \nodata &  5 \\
\enddata

\tablerefs{
(1)  This work;
(2)  SDSS DR7
(3)  \citet{BFT2009};
(4)  \citet{GSM2009};
(5)  \citet{ATN2009}
}

\end{deluxetable}

\subsection{Luminosity and Star-Formation Rate\label{SECTION:host_lum}}

     SDSS J135846.65+470104.5 has a colour of $B\!-\!R = +1.91 \pm
0.13$, making it a fairly red galaxy.  The typical absolute magnitude
for a red galaxy at $0.2 \le z < 0.5$ is ${(M^*_B)}_\mathrm{AB} =
-20.44$ \citep{LTF1995}, assuming a cosmology with
$(H_0,\Omega_m,\Omega_\Lambda) = (50,1,0)$.  For our adopted
cosmology that corresponds to ${(M^*_B)}_\mathrm{AB} = -20$.  At $z =
0.345$ the rest-frame $B$ band corresponds roughly to the observed
$r^\prime$ band.  Assuming that the galaxy has a power-law spectrum
(see \S~\ref{SECTION:sed}) and $r^\prime = 21.6$, then it has a
rest-frame luminosity in the $B$ band of $L_B \approx 1.3 L^*_B$ where
$L^*_B$ is the rest-frame $B$-band luminosity of a typical red galaxy
at $z = 0.345$.  This indicates that the host is approximately an
$L^*$ galaxy with a total luminosity that is roughly similar to that
of the Milky Way.  However, we stress that this result is somewhat
uncertain because $M^*$ is highly correlated with the slope of the
faint end of the galaxy luminosity function, and with its
normalization \citep{LTF1995}.

We can estimate the star-formation rate in the host galaxy using Eq.~2
of \citet{MPD1998}.  At the redshift of the burst 1500~\AA\
approximately corresponds to the UVOT uvw2 filter while 2800~\AA\
approximately corresponds to the SDSS $u^\prime$ filter.  Correcting
for Galactic extinction these magnitudes for the host become uvw2$_0 =
22.55$ and $u^\prime_0 = 23.15$.  These yield star-formation rates (in
$\mathcal{M}_\Sun$ yr$^{-1}$) of 0.3 (1500~\AA) and 0.7 (2800~\AA)
assuming a \citet{S1955} initial mass function, and 0.6 (1500~\AA) or
1.2 (2800~\AA) assuming a \citet{S1986} initial mass function.  These
values assume that there is no extinction in the host galaxy.  This is
clearly not correct, so they actually represent a lower limit on the
star-formation rate of approximately 1~$\mathcal{M}_\Sun$ yr$^{-1}$.
Correcting for the mean extinction in the host ($A_v = 3.5$ mag) gives
a star-formation rate of $\ga 10^2$~$\mathcal{M}_\Sun$ yr$^{-1}$.
However, this corrected values should be treated with caution since
extinction in the host is likely to be highly variable and may not be
affecting all star-forming regions equally.

     The host galaxy appears to be roughly similar in luminosity 
to the Milky Way, but with a significantly higher star-formation rate.
This puts it on the massive end of the GRB host galaxy distribution.


\section{The Gamma-Ray Burst and its Afterglow\label{SECTION:grb}}

\subsection{Is GRB~090417B a Dark Burst?\label{SECTION:dark}}

No afterglow has been detected for GRB~090417B at infrared, optical,
or ultraviolet wavelengths.  Changes in the near-ultraviolet flux from
the host galaxy between early and late times are consistent with no
contribution from an afterglow down to a limiting magnitude of
$u_\mathrm{oa} \ga 24.9$.  We are assuming that the bright limit on
the afterglow magnitude is valid for all time after four hours after
the BAT trigger.  There is no evidence for a brightening of the host
galaxy at late times, so this is a reasonable assumption.  Therefore
we adopt this as an upper limit to the luminosity of the afterglow at
11~hr and compare it to the $X$-ray flux at 11~hr.  This yields
$\beta_\mathrm{OX} \le -1.9$, making it a dark burst \citep{JHF2004}.
The $X$-ray spectrum at this time is evolving from $\beta_X = 1.3$ to
$\beta_X = 2.3$, so using the definition of \citet{HKG2009} we find
that $\beta_\mathrm{OX} < \beta_\mathrm{X} - 0.5$, also making GRB
090417B a dark burst.  Our limit on $\beta_\mathrm{OX}$ is somewhat
uncertain due to the large uncertainty in our estimate of
$u_\mathrm{oa}$.  However, for GRB~090417B to be an optically bright
burst by the definition of \citep{JHF2004} then the optical afterglow
would have had to have had $u \la 21.7$, which is inconsistent with
the observed host+afterglow magnitude ($u = 23.09 \pm 0.38$).

     Our computation of the spectral index between the optical and
$X$-ray regimes depends on the limiting magnitude of the afterglow
in the UVOT $u$ band as determined from the statistical error in the
observed magnitude of the host galaxy at two epochs.  In order to test
that GRB~090417B really is a dark burst we compute the expected UVOT
$u$-band magnitude for the afterglow given the observed $X$-ray
spectrum and assuming that the cooling frequency is between $X$-ray
and optical wavelengths.  This yields a prediction of $u \la 21.7$ at
11 hours, which is ruled out by the observations.  Therefore, we
conclude that GRB~090417B is a dark GRB\@.

     The difference between the predicted $u$ magnitude and the upper
limit that we derive in \S~\ref{SECTION:uvot_data} is $\Delta u \ga
3.2$ mag.  If we assume a Milky Way extinction law this corresponds to
$A_V \ga 2.5$ mag of extinction in the host galaxy along the line of
sight to the burst.

\subsection{Closure Relations\label{SECTION:closure}}

     If we assume the canonical $X$-ray light curve for GRB afterglows
\citep{NKG2006,ZFD2006} then we can use the temporal decay indices and
the spectral slopes to constrain the underlying physics.  We stress
that the following analysis is only valid if we assume the
relativistic fireball model.  Alternate models for the $X$-ray
afterglows of GRBs are available, such as
the cannonball model \citep{DD2000a,DD2000b,DDD2002a,DDD2002b},
the disc model of \citet{CG2009},
the late internal emission model \citep{GGN2007,KNJ2008},
the long-lived reverse shock model \citep{GDM2007,UB2007}, and
the prior outflow emission model \citep{Y2009,LLZ2009}.

The time and post-break decay index of the late-time break in the
$X$-ray light curve at $4.5^{+2.8}_{-1.2} \times 10^5$~s after the BAT
trigger are consistent with a jet break.  However, the dust scattering
interpretation of the late-time spectral softening (see
\S~\ref{SECTION:dust_scattering}) suggests that the steepening of the
light curve after $\approx 4.5 \times 10^5$~s is due to scattered
emission due to dust dominating over the synchrotron component of the
afterglow.  Unfortunately there is no data in other wavelength regimes
to test if the late-time steepening is achromatic (which would argue
for a jet break), therefore it is uncertain if this is a jet break.

     \citet{SPH1999} give the relationships between the temporal and
spectral slopes and the electron power-law distribution index, $p$,
for a blast expanding into a homogeneous external medium, and
\citet{CL1999} give the relationships for a stellar wind ($\rho(r)
\propto r^{-2}$) environment.  Both of these papers assume that $p >
2$.  For the case were $p < 2$ the relationships of \citet{DC2001} are
used.  \citet{PK2002} found that GRB afterglows exhibit a range of
electron distributions with values for the ten GRBs they studied being
between $p = 1.4$ and $p = 2.8$.

     The $X$-ray spectral index during the first part of phase III of
the afterglow ($9100 \le t \le 29\,600$~s) is $\beta_X = \Gamma - 1 =
1.3 \pm 0.2$.  If the cooling frequency is above the $X$-ray
regime during this time then the predicted electron index is $p =
2\beta_X + 1 = 2(1.3 \pm 0.2) + 1 = 3.6 \pm 0.4$, which is well
outside the range of values found by \citet{PK2002}.  If the cooling
frequency is below the $X$-ray band during this time then $p =
2\beta_X = 2(1.3 \pm 0.2) = 2.6 \pm 0.4$, which is consistent with the
\citet{PK2002} range of electron indices.  For $\nu_c < \nu_X$ the
spectral index predicts a temporal decay of $\alpha_X = 3/2\beta_X -
1/2 = 3/2(1.3 \pm 0.2) - 1/2 = 1.45 \pm 0.30$ for either a homogeneous
interstellar medium (ISM) or a stellar wind environment.  The
observed $X$-ray decay during this time is $\alpha_X = 1.40 \pm 0.04$,
which is consistent with the predicted value.  Therefore, we conclude
that the cooling frequency is most likely below the $X$-ray band at
$9100 \le t \le 29\,600$~s.  Unfortunately it is not possible to tell,
from the $X$-ray data, if the burst is expanding into a uniform
external medium or a pre-existing stellar wind.  The broad agreement
between $\alpha_2$ and $\beta_2$ with the predictions of the fireball
model suggests that this is a reasonable explanation for the physics
of this afterglow during this phase of the afterglow.

     Between 68\,980~s and 428\,400~s the spectral index softens to
$2.8 \pm 0.5$.  In this interval the fireball model predicts $\alpha_X
= 3.7 \pm 1.0$ for a homogeneous ISM and a wind if $\nu_c < \nu_X$.
If $\nu_c > \nu_X$ then the environment must be windy (since $\nu_c$
is increasing) and $\alpha_X$ is predicted to be $4.7 \pm 1.0$.  The
observed value is $\alpha_2 = 1.40 \pm 0.04$.  Neither case is
consistent with the data in this time interval.

     The data before 68\,980~s is broadly consistent with a
relativistic fireball expanding into either a homogeneous ISM or a
wind-stratified external medium, and a cooling frequency below the
$X$-ray band in the period $9\,100 \le t \le 29\,600$~s.  However,
this model can not explain the data later than 68\,980~s.  Some other
physical process is needed to explain the late-time softening of the
$X$-ray spectrum.


\subsection{Dust Scattering\label{SECTION:dust_scattering}}

     For GRBs occurring in dusty star forming regions, the $X$-rays
emitted during the prompt phase may be scattered by dust grains in
small angles, which leads to a delayed $X$-ray afterglow component
\citep{MG2000}.  Placing circumburst dust at $\approx$ 10--100 pc away
from the burst, the calculated dust echo $X$-ray lightcurve displays a
shallow decay lightcurve followed by late steepening, as is commonly
observed in many GRBs \citep{SD2007,SDM2008}.  One major feature of
this model is the strong spectral softening evolution in the echo
emission \citep{SWK2009}.  Such a softening feature has not commonly
been observed in the majority of GRBs, so the dust scattering model is
not favored to interpret those bursts.  However, GRB 090417B shows a
spectral softening during the shallow decay phase.  Also the
optically dark nature of the afterglow suggests that there is a
substantial amount of dust in the circumburst environment.  All these
make this burst a very probable candidate for the dust scattering
model.  In modeling this burst, we assume a dust sheet with distance
$R_d$ from the burst, with other dust grain properties the same as in
Figure~1 of \citet{SWK2009}.  We find the calculated light curve can
reproduce quite well the XRT light curve from $T-T_0= 5\times 10^3$~s
to $10^5$~s (we consider the flares prior to this time as not being
due to dust scattering because the $\gamma$-ray emission was still on
at that time) for $R_d \approx$ 30--80~pc and for a broad range of the
source emission spectral index values.

\begin{figure}

  \includegraphics[angle=+270]{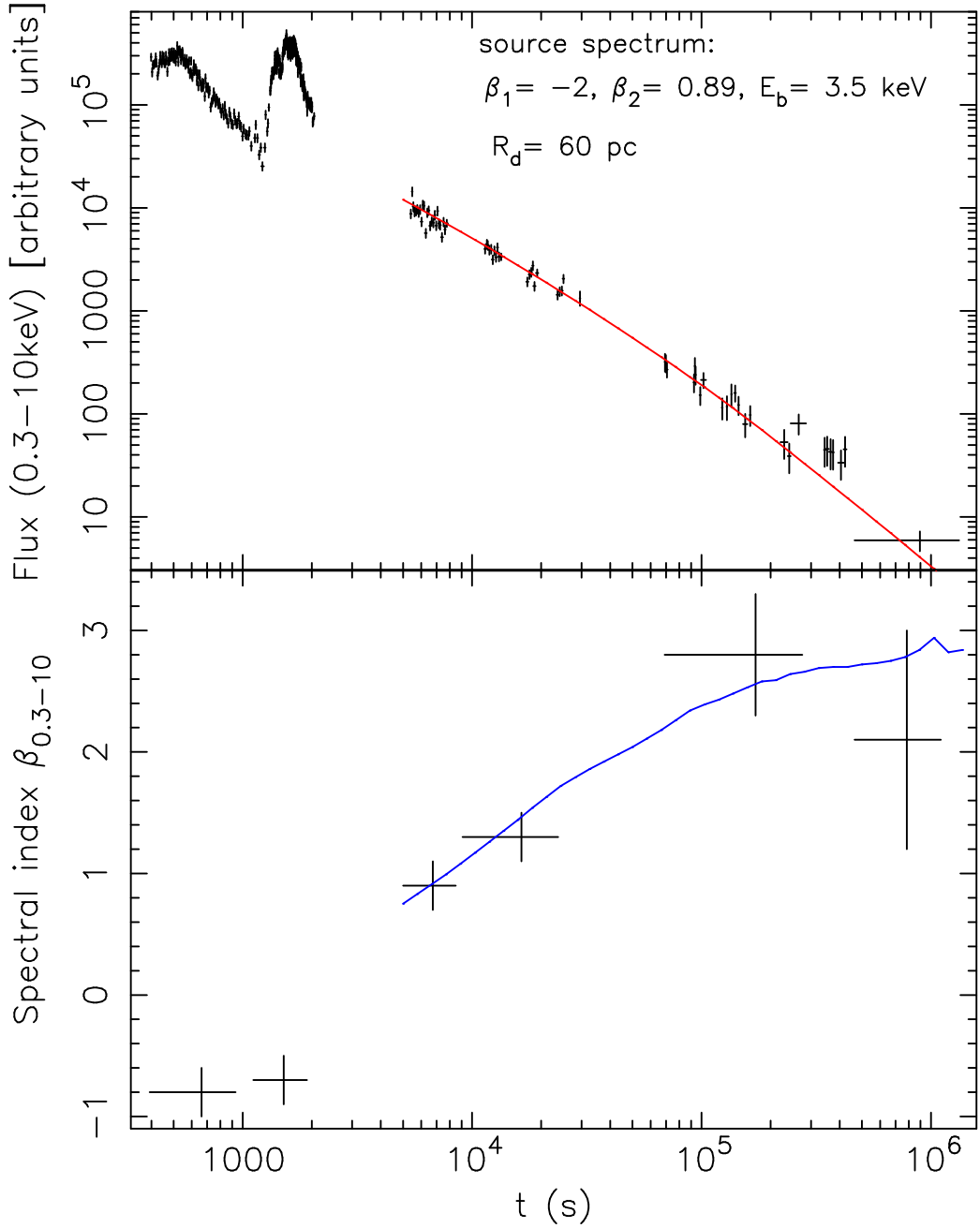}

\figcaption[./fit_R_60_Eb_3.5.ps]{Dust scattering model fit to
  GRB~090417B $X$-ray data. Top panel: the fit to the light
  curve. Bottom panel: the fit to the spectral index evolution.  The
  model predictions are shown as solid lines and data as crosses.  The
  model parameter values are same for both panels, and are as labeled.
  The source spectrum is modeled as a two-piece power law: $f_{\nu}
  \propto \nu^{-\beta_1}$ and $f_{\nu} \propto \nu^{-\beta_2}$ with a
  break $E_b$.  In the fit, the low energy spectral index $\beta_1=
  -2$ is adopted, suggesting a self-absorbed prompt emission
  spectrum. $\beta_2$ is adopted to be equal to the reported prompt
  $\gamma$-ray spectral index.\label{FIGURE:dust_fit}}
\end{figure}

A strong test of the model is whether the spectral index evolution can
be reproduced.  In Figure~\ref{FIGURE:dust_fit} we plot the temporal
evolution of the observed XRT spectral index.  The dust scattered
emission spectral index is calculated in the same way as in
\citet{SWK2009}.  The amount of softening predicted by the model is
large, {\ie}, $\Delta \beta \approx$ 3--4, and it is independent of
the source (prompt) emission spectral index.  We tried to use the
reported prompt BAT spectral index, $\beta_{BAT}= 0.89$, and the even
harder spectral indices observed in the two $X$-ray flares ($T-T_0=
400 - 2\times10^3$~s), $\beta_X= -1$, as the source $X$-ray spectral
index, but found that the model predicted emission is significantly
softer than what is observed.  However, if one assumes an even harder
spectral index of $\beta_X = -2$ for the prompt $X$-rays, both the
$X$-ray light curve and spectral evolution can be properly reproduced
(see Figure~\ref{FIGURE:dust_fit}).  Such a hard $X$-ray spectrum is
consistent with the self-absorbed regime of a synchrotron or inverse
Compton scattering spectrum.  We therefore assume that the GRB prompt
emission spectrum has a break above the $X$-ray band due to the
self-absorption frequency, $\nu_a$, below which the spectrum has
$\beta = -2$ or $-5/2$ \citep[\eg,][]{GS2002}. A large $\nu_a$ is
possible if the prompt emission radius is small enough
\citep[\eg,][]{SZ2009}.  The best model fit to the lightcurve and the
spectral evolution is shown in Figure~\ref{FIGURE:dust_fit}.  The fit
to the $X$-ray light curve has a reduced $\chi^2$ of 2.7, and the fit
to the spectral index evolution has a reduced $\chi^2$ of 1.5.

We also estimate the amount of dust in terms of $\tau_0$, the
scattering optical depth to the 1~keV photons, using the same formula
as in \citet{SWK2009}.  The source $X$-ray fluence is extrapolated
from the observed $\gamma$-ray fluence using the assumed spectral
form, and the scattered emission $X$-ray fluence is estimated from the
observed XRT fluence for $T-T_0 \geq 5\times10^3$~s using an
intermediate value of observed spectral index $\beta_X= 1.5$.  We find
$\tau_0 \approx 2.5$.  This means that over half of prompt 1~keV
photons would be scattered by the dust.  This $\tau_0$ value also
implies $A_V \approx 15 - 40$~mag, according to some empirical
relations between the two dust properties used in \citet{SWK2009}.
This $A_V$ is much higher than the average value for the host galaxy
that was derived from the spectral energy distribution in
Section~\ref{SECTION:sed}, but is consistent with the extinction
derived from the $X$-ray measurement of the hydrogen column along the
line of sight to the burst, $A_V \approx 11$~mag.  This high
line-of-sight extinction naturally explains the non-detection of an
optical afterglow for GRB~090417B.  Our model indicates that the dust
grain column density along the line of sight is $\approx 10^{12}$
cm$^{-2}$, and the density of the intervening dust sheet is $n \ga
10^{-8}$ cm$^{-3}$.

The data of GRB~090417B is consistent with the dust scattering model
under the condition that the prompt $X$-rays have a self-absorbed
spectrum in the $X$-ray range.  The large amount of dust inferred from
the $X$-ray modeling is consistent with the dark nature of the optical
afterglow.


\subsection{Energetics of the Burst\label{SECTION:energetics}}

BAT observations of GRB~090417B were cut off at approximately 2105~s
after the BAT trigger when {\sl Swift\/} went into an Earth limb
constraint.  Therefore the total duration, and thus the total fluence,
of the burst is not known.  However, the fluence while the burst was
observed was $8.20^{+1.0}_{-2.1} \times 10^{-6}$ erg cm$^{-2}$ in the
15--150~keV band.  This puts a lower limit on the total fluence.  We
can use this lower limit to estimate a lower limit on the isotropic
equivalent energy of GRB~090417B.
The BAT spectrum is best fit by a single power law, so at a redshift
of $z = 0.345$ this corresponds to an isotropic equivalent energy of
$E_\mathrm{iso} > 2.4^{+0.3}_{-0.6} \times 10^{51}$~erg.  We performed
a $k$ correction following the prescription of \citet{BFS2001} and
find $k = 2.6$ (with some uncertainty due to the arbitrary choice of
cut-off energy for the power-law spectrum).  This yields a
$k$-corrected total isotropic energy of $E_\mathrm{iso} \ga 6.3 \times
10^{51}$~erg in the 20--2000 keV band.  Our data does not allow us to
constrain the time of a jet break, so we are unable to estimate the
jet opening angle or the intrinsic gamma-ray energy of GRB~090417B.


\subsection{Constraints on a Supernova Component\label{SECTION:sn}}

To date no supernova signature has been seen for any dark GRB\@.  If a
supernova were detected in a dark GRB that would have profound
implications for origin of that dark burst.  The detection of a
supernova would indicate that the burst was not dark due to extinction
or high redshift, but must have a fundamentally different afterglow
than conventional optically-bright bursts.  In light of this we
examined our data to see if there is any evidence for a supernova
component for GRB~090417B.

UVOT observations were taken up to 15 days after the BAT trigger, so
we searched this late-time data for evidence of a supernova component.
A type Ib/c supernova like SN1998bw \citep{PP1998} at a redshift of $z
= 0.345$ is expected to peak at approximately $10(1+z) \approx 13$
days after the burst.  We find no evidence for a change in the
luminosity of the host galaxy down to $\Delta u \approx 0.2$~mag,
which corresponds to a upper limit on the magnitude of the supernova
of $u_{\lim} \ga 24.9$.  At $z = 0.345$ the central wavelength of the
UVOT $u$ band corresponds to $\lambda_0 \approx 2600$~\AA\@.  No
observational data is available for the peak magnitude of SN1998bw at
2600~\AA\@.  However {\sl Swift\/}/UVOT ultraviolet light curves are
available for SN2007Y \citep{BHI2009}, a SN~Ib/c with a peak absolute
magnitude of uvw1 $\approx -17$ in the rest frame.  This corresponds
to a rest frame uvw1 of $\approx 24$ at the redshift of GRB 090417B.
This is similar to our limiting magnitude, so if we assume that
SN2007Y is typical of the SNe~Ib/c associated with GRBs we are not
able to constrain the existence of a supernova associated with GRB
090417B.


\section{Discussion\label{SECTION:discussion}}

     All of the evidence presented in this paper points to GRB~090417B
being a dark burst because of localized extinction in the host galaxy.
The host galaxy has a redshift of $z = 0.345$, and is clearly seen at
optical and ultraviolet wavelengths.  Therefore Lyman-$\alpha$
absorption can not account for the missing optical afterglow.  The
$X$-ray light curve and spectrum during the first $\approx 70$~ks obey
the conventional closure relationships and are consistent with
synchrotron radiation from a relativistic fireball expanding into an
external medium (although the structure of that medium can not be
determined with the available data).  Optical afterglows have been
detected for many GRBs with similar $X$-ray properties, so it unlikely
that the lack of an optical detection is due to unusual physics during
the afterglow phase of the burst.  Optical observations of GRB~090417B
started 378~s after the BAT trigger, and follow-up observations were
made from several observatories for several days after the burst.
These observations went deep enough to have detected an optical
afterglow if one had been present with a luminosity similar to that of
other optical afterglows.  Therefore, the dark nature of GRB~090417B
can not be explained by a lack of observations.

     The dust hypothesis, however, can explain this dark burst.  The
$X$-ray spectrum is consistent with an \ion{H}{1} column density of
$N_\mathrm{H} \approx 2 \times 10^{22}$~cm$^{-2}$, which corresponds
to an extinction of $A_v \approx 11$~mag along the line of site in the
host galaxy, assuming a Milky Way extinction law as explained in
\S~\ref{SECTION:xrt_data}.  This value is in agreement with the
minimum extinction, as explained in \S~\ref{SECTION:dark}, that is
needed to obscure the expected bright optical afterglow ($A_V \ga
2.5$~mag).  An extremely high extinction ($A_V \approx 15$--40~mag) is
also predicted by the dust scattering model, which is consistent with
the $X$-ray data.  Such a large amount of dust along the line of sight
to the GRB naturally explains the lack of any detection at
ultraviolet, optical, or near-infrared wavelengths while allowing a
conventional $X$-ray afterglow to be observed.  It is also consistent
with the result of \citet{PCB2009} who found large mean extinctions for
the putative host galaxies of several GRBs.  Our result, that the
extinction along the line of sight in the host GRB~090417B is likely
to be at least ten magnitudes in the UVOT $v$ band, provides strong
evidence that local dust concentrations in the host galaxy are
responsible for at least some low-redshift dark bursts.

     Unlike many GRB host galaxies the dust in the host galaxy of
GRB~090417B appears to follow a Milky Way extinction law.  In addition
the host appears to be an $L^*$ galaxy with a star-formation rate that
exceed $\approx 1$ $\mathcal{M}_\sun$ yr$^{-1}$.  In other words, the
host appears to have dust properties, and a luminosity that are
similar to those of the Milky Way, and a star-formation rate that is
consistent with what is seen in other GRB host galaxies.

GRB~090417B is one of the few dark GRBs where the association with a
host galaxy is secure.  In most cases associations between a GRB and a
galaxy have been made based on the probability of the nearest observed
galaxy lying as close to the centre of the $X$-ray error circle as it
does.  There are a few exceptions, such as GRB~000210, GRB~050713A and
XRF~050416A.  The host galaxy of GRB~000210 was identified from the
locations of its radio and $X$-ray afterglow.  \citet{PFG2002} found a
probability of a chance alignment of the galaxy and the burst of
$P_\mathrm{ch} = 0.016$ and conclude that if this is the host then the
dark nature of GRB~000210 is likely due to dust either at the location
of the progenitor or along the line of sight.  XRF~050416A occurred at
a low redshift ($z = 0.6535$ \citet{CKG2005}).  Although technically
dark an optical afterglow was detected and was consistent with the
standard fireball model \citep[\eg,][]{CF2005,HBG2007}.  The presence
of an optical afterglow and the low line-of-sight extinction suggests
that this burst should not be considered to be a dark GRB\@.  The
line-of-sight extinction for XRF~050416A is $A_V < 1$~mag
\citep{HBG2007,PCB2009}, which is consistent with what is seen for
most optically detected GRB afterglows.  \citet{PCB2009} note that
GRB~050713A also only barely qualifies as a dark burst under the
\citet{JHF2004} definition.  The internal line-of-sight extinction for
this burst is unknown.  However, an optical afterglow was seen
\citep{WVW2005}, so it is unlikely that the extinction was large.  The
latter two of these bursts are technically dark based on the spectral
slope between the $X$-ray and optical regimes ($\beta_{OX}$) criteria
of \citet{JHF2004}.  However, in these cases optical or infrared
afterglows were detected, suggesting that these GRBs were borderline
cases of dark bursts.

Further evidence that dust is responsible for GRB~090417B being dark
comes from the late-time softening of the $X$-ray spectrum.  This
softening is consistent with what is predicted from the dust
scattering model \citep{SD2007,SDM2008,SWK2009}, and requires
approximately 15--40~mag of extinction at optical wavelengths.  The
effects of dust scattering do not become apparent in the $X$-ray
spectrum until $\approx 10^5$~s after the burst.  Since $X$-ray
spectra are usually extracted from early-time data (when there are a
large number of photons) it is possible that the softening seen for
GRB~090417B has been present in the late-time $X$-ray spectra of other
dark bursts, but has not been seen due to low count rates.  For
example, the standard catalogue of {\sl Swift\/}/XRT spectral fits
\citep{EBP2009} uses data taken within $\approx 4.3$~ks of the BAT
trigger, well before a softening due to dust scattering is expected to
appear.

\citet{PCB2009} point out that the host galaxies of dark GRBs are not
unusually reddened relative to the hosts of optically bright GRBs.
Their analysis assumes that their statistical association of bursts
with the nearest detected galaxy on the sky reveals the true hosts of
these bursts.  It is uncertain if this is valid for all of the GRBs in
their sample.  However, it is likely that at least some (if not all)
of these galaxies actually did host the associated GRBs.  Although
some of the hosts of dark GRBs have been found to be unusually dusty
compared to the hosts of optically-bright GRBs ({\eg}, GRB~030115
\citep{LFR2006}), there is no evidence that the hosts of dark GRBs
have systematically higher mean extinctions than the hosts of
optically bright GRBs.  However, the mean extinction of the host is
probably not the factor that determines if a particular burst is dark
or not since any extinction due to material that is not along the line
of sight to the burst will not affect our observations of the
afterglow.  What is important in making a burst optically dark or
bright is the extinction along the line of sight.  This extinction can
be due to dust at the location of the GRB, although it has been
suggested that the ultraviolet and $X$-ray emission from GRBs can
destroy circumburst dust out to a few tens of pc from the progenitor
\citep{WD2000,FKR2001}.  One way to get a dusty environment around a
GRB progenitor is the stellar wind from a Wolf-Rayet progenitor.
Collisions between wind-driven shells of material ejected during
various stages of the Wolf-Rayet star's evolution can trigger the
formation of large amounts of dust
\citep[\eg][]{WVT1990,SDM2008,C2009}.  The radius of a Wolf-Rayet wind
bubble is consistent with with the distance from the progenitor that
we find for the dust along the line of sight to GRB~090417B (tens of
pc).  Alternately, the extinction could be due to unrelated dust in
the host that just happens to lie along the line of sight, such as a
dusty star-forming region or a giant molecular cloud.  Giant molecular
clouds have diameters of approximately 100 pc \citep{MYM2001}, but
often contain dense cores where star formation occurs.  A chance
alignment of one of these cores along the line of sight to GRB 090417B
could account for the high extinction inferred from the X-ray data.

For GRB~090417B the late-time evolution of the $X$-ray spectrum can be
explained if there is a sheet of dust at a distance of $30 \la R_d \la
80$~pc from the burst.  This suggests that GRB~090417B may have been
dark because of either a dusty environment within 30--80 pc of the
progenitor \citep[\eg][]{WD2000,FKR2001}, or a chance alignment of the
progenitor with a region of high extinction in the host galaxy.  The
Galactic extinction model of \citet{DCL2003} suggests that roughly
25\% of the sight lines through the Milky Way have $A_V \ga 1$~mag.
The modelling of \citet{UHG2009} suggests that most sightline through
a galaxy like the Milky Way, as seen from within the galaxy, have $A_V
\la 1$~mag.  However, their work does not include clumpiness in the
dust distribution, which would lead to a larger fraction of sightlines
having higher extinctions.  This suggests that a non-negligible
(albeit uncertain) fraction of GRBs may be located along a sightline
with more than approximately one magnitude of extinction in their
hosts.  When high-redshift dark bursts are accounted for the fraction
of dark bursts is roughly comparable of the estimated fraction of
highly extincted sightlines, which suggests that obscuration by dust
along the line of sight may be responsible for some dark GRBs.


\section{Conclusions\label{SECTION:conc}}

We have shown that the host of the dark GRB~090417B is very likely to
be SDSS~J135846.65+470104.5, an $L^*$ galaxy at $z = 0.345$.  This
galaxy has an overall star-formation rate that is at least as great as
that of the Milky Way, and an overall $V$-band extinction of $A_V
\approx 3.5$ mag, which is dustier than typical GRB host galaxies, but
not greatly different from the Milky Way.

     $X$-ray observations of GRB~090417B show the normal temporal and
spectral behaviour seen in the $X$-ray afterglows of many {\sl
  Swift\/} GRBs.  The data up to approximately 70~ks after the BAT
trigger are consistent with a relativistic fireball expanding into
either a homogeneous ISM or a wind-stratified external medium.  The
cooling frequency is below the $X$-ray band between 9.1~ks and
29.6~ks, and the $k$-corrected isotropic energy is $E_\mathrm{iso} \ga
6.3 \times 10^{51}$ erg.
The physics of this burst do not appear to be unusual, and thus are
unlikely to explain the dark nature of the burst.

     After approximately 70~ks the $X$-ray spectrum becomes
significantly softer.  We find that this can be explained using the
dust scattering model of \citet{SD2007}.  The observed late-time
spectral evolution can be produced by a sheet of dust approximately
30--80 pc from the burst along the line of sight.  The model predicts
an extinction of $A_V \approx 15$--40~mag along the line of sight.
This is consistent with the \ion{H}{1} column density measured from
the $X$-ray spectrum.  Therefore, we conclude that GRB~090417B is
probably dark because of a dense, localized layer of dust along the
line of sight between us and the afterglow.


\acknowledgements

We acknowledge the use of public data from the {\sl Swift\/} Data
Archive.  This paper is based in part on observations taken with the
Nordic Optical Telescope, operated on the island of Santa Miguel de la
Palma jointly by Denmark, Finland, Iceland, Norway, and Sweden in the
Spanish Observatorio del Roque de los Muchachos of the Instituto de
Astrof{\'\i}sica de Canarias.  The authors would like to thank the
anonymous referee for their comments, which improved this paper.


  

\begin{thebibliography}{}

\bibitem[Aoki {\etal}(2009)]{ATN2009}
  Aoki, K.,
  Tanaka, I.,
  Nakata, F.,
  Ohta, K.,
  Yuma, S., \&
  Kawai, N.,
  2009, GCNC 9145

\bibitem[Barthelmy {\etal}(2005)]{BBC2005}
  Barthelmy, S.~D.,
  Barbier, L.~M.,
  Cummings, J.~R.,
  Fenimore, E.~E.,
  Gehrels, N.,
  Hullinger, D.,
  Krimm, H.~A.,
  Markwardt, C.~B.,
  {\etal},
  2005, Sp.\ Sci.\ Rev., 120, 143

\bibitem[Berger \& Fox(2009)]{BF2009}
  Berger, E., \&
  Fox, D.~B.,
  2009, GCNC 9156

\bibitem[Berger {\etal}(2007)]{BFK2007}
  Berger, E.,
  Fox, D.~B.,
  Kulkarni, S.~R.,
  Frail, D.~A., \&
  Djorgovski, S.~G.,
  2007, \apj, 660, 504

\bibitem[Berger {\etal}(2009)]{BFT2009}
  Berger, E.,
  Fox, D.~B., \&
  Tanvir, N.,
  2009, GCNC 9158

\bibitem[Bertin \& Arnouts(1996)]{BA1996}
  Bertin, E., \&
  Arnouts, S.,
  1996, \aaps, 117, 393

\bibitem[Bloom {\etal}(2001)]{BFS2001}
  Bloom, J.~S.,
  Frail, D.~A., \&
  Sari, R.,
  2001, \aj, 121, 2879

\bibitem[Breeveld {\etal}(2010)]{BCH2010}
  Breeveld, A.~A.,
  Curran, P.~A.,
  Hoversten, E.~A.,
  Koch, S.,
  Landsman, W.,
  Marshall, F.~E.,
  Page, M.~J.,
  Poole, T.~S.,
  {\etal},
  2010, \mnras, in press, arXiv:1004.2448

\bibitem[Bloom {\etal}(2002)]{BKD2002}
  Bloom, J.~S.,
  Kulkarni, S.~R., \&
  Djorgovski, S.~G.,
  2002, \aj, 123, 1111

\bibitem[Brown {\etal}(2009)]{BHI2009}
  Brown, P.~J.,
  Holland, S.~T.,
  Immler, S.,
  Milne, P.,
  Roming, P.~W.~A.,
  Gehrels, N.,
  Nousek, J.,
  Panagia, N.,
  {\etal},
  2009, \aj, 137, 4517

\bibitem[Burrows {\etal}(2005)]{BHN2005}
  Burrows, D.~N.,
  Hill, J.~E.,
  Nousek, J.~A.,
  Kennea, J.~A.,
  Wells, A.~A.,
  Osborne, J.~P.,
  Abbey, A.~F.,
  Beardmore, A.,
  {\etal},
  2005, Sp.\ Sci.\ Rev.\, 120, 165 

\bibitem[Campana {\etal}(2006)]{CMB2006}
  Campana, S.,
  Mangano, V.,
  Blustin, A.~J.,
  Brown, P.,
  Burrows, D.~N.,
  Chincarini, G.,
  Cummings, J.~R.,
  Cusumano, G.,
  {\etal},
  2006, \nat, 442, 1008

\bibitem[Cannizzo \& Gehrels(2009)]{CG2009}
  Cannizzo, J.~K., \&
  Gehrels, N.,
  2009, \apj, 700, 1047

\bibitem[Cash(1979)]{C1979}
  Cash, W.,
  1979, \apj, 228, 939

\bibitem[Cenko \& Fox(2005)]{CF2005}
  Cenko, S.~B., \&
  Fox, D~.B.,
  2005, GCNC 3265

\bibitem[Cenko {\etal}(2005)]{CKG2005}
  Cenko, S.~B.,
  Kulkarni, S.~R.,
  Gal-Yam, A., \&
  Berger, E.,
  2005, GCNC 3542

\bibitem[Cherchneff(2009)]{C2009}
  Cherchneff, I.,
  2009, in ASP Conf.\ Ser.,
  Hot and Cool: Bridging Gaps in Massive Star Evolution,
  eds. C. Leitherer, P.~D., Bennett, P.~W. Morris, \& J.~T. van~Loon,
  arXiv:0909.0164

\bibitem[Chevalier \& Li(1999)]{CL1999}
  Chevalier, R.~A., \&
  Li, Z.-Y.,
  1999, \apjl, 520, L29

\bibitem[Christensen {\etal}(2004)]{CHG2004}
  Christensen, L.,
  Hjorth, J., \&
  Gorosabel, J.,
  2004, \aap, 425, 913

\bibitem[Conselice {\etal}(2005)]{CVF2005}
  Conselice, C.~J.,
  Vreeswijk, P.~M.,
  Fruchter, A.~S.,
  Levan, A.,
  Kouveliotou, C.,
  Fynbo, J.~P.~U.,
  Gorosabel, J.,
  Tanvir, N.~R.,
  {\etal},
  2005, \apj, 633, 29

\bibitem[Cummings {\etal}(2009)]{CBB2009}
  Cummings, J.~R.,
  Barthelmy, S.~D.,
  Baumgartner, W.~H.,
  Fenimore, E.~E.,
  Gehrels, N.,
  Krimm, H.~A.,
  Markwardt, C.~B.,
  Palmer, D.~M.,
  {\etal},
  2009, GCNC 9139
  
\bibitem[Dado {\etal}(2002a)]{DDD2002a}
  Dado, S.,
  Dar, A., \&
  De~R{\'u}jula, A.,
  2002a, \aap, 388, 1079

\bibitem[Dado {\etal}(2002b)]{DDD2002b}
  Dado, S.,
  Dar, A., \&
  De~R{\'u}jula, A.,
  2002b, AAp, 401, 243

\bibitem[Dai \& Cheng(2001)]{DC2001}
  Dai, Z.~G., \&
  Cheng, K.~S.,
  2001, \apjl, 558, L109

\bibitem[Dar \& De~R{\'u}jula(2000a)]{DD2000a}
  Dar, A., \&
  De~R{\'u}jula, A.,
  2000a, astro-ph/0008474

\bibitem[Dar \& De~R{\'u}jula(2000b)]{DD2000b}
  Dar, A., \&
  De~R{\'u}jula, A.,
  2000b, astro-ph/0012227

\bibitem[de~Vaucouleurs(1948)]{dV1948}
  de~Vaucouleurs, G.,
  1948, AnAp, 11, 247
  
\bibitem[de~Vaucouleurs(1959)]{dV1959}
  de~Vaucouleurs, G.,
  1959, HDP, 53, 275 \& 311

\bibitem[Drimmel {\etal}(2003)]{DCL2003}
  Drimmel, R.,
  Cabrera-Lavers, A., \&
  L{\'o}pez-Corredoira, M.,
  2003, \aap, 409, 205
  
\bibitem[El{\'\i}asd{\'o}ttir {\etal}(2009)]{EFH2009}
  El{\'\i}asd{\'o}ttir, {\'A}.,
  Fynbo, J.~P.~U.,
  Hjorth, J.,
  Ledoux, C.,
  Watson, D.~J.,
  Andersen, A.~C.,
  Malesani, D.,
  Vreeswijk, P.~M.,
  {\etal},
  2009, ApJ, 697, 1725

\bibitem[Evans {\etal}(2009)]{EBP2009}
  Evans, P.~A.,
  Beardmore, A.~P.,
  Page, K.~L.,
  Osborne, J.~P.,
  O'Brien, P.~T.,
  Willingdale, R.,
  Starling, R.~L.~C.,
  Burrows, D.~N.,
  {\etal},
  2009, \mnras, 397, 1177

\bibitem[Fishman {\etal}(1988)]{FM1989}
  Fishman, G.~J.,
  Meegan, C.~A.,
  {\etal},
  1989, Proc.\ GRO Science Workshop, GSFC, 2

\bibitem[Fruchter {\etal}(2001)]{FKR2001}
  Fruchter, A.~S.,
  Krolik, J.~H., \&
  Rhoads, J.~E.
  2001, ApJ, 563

\bibitem[Fruchter {\etal}(2006)]{FLS2006}
  Fruchter, A.~S.,
  Levan, A.~J.,
  Strolger, L.,
  Vreeswijk, P.~M.,
  Thorsett, S.~E.,
  Bersier, D.,
  Burud, I.,
  Castro~Cer{\'o}n, A.~J.,
  {\etal},
  2006, \nat, 441, 463

\bibitem[Fynbo {\etal}(2001)]{FJG2001}
  Fynbo, J.~P.~U.,
  Jensen, B.~L.,
  Gorosabel, J.,
  Hjorth, J.,
  Pedersen, H.,
  M{\o}ller, P.,
  Abbott, T.,
  Castro-Tirado, A.~J.,
  {\etal},
  2001, \aap, 369, 373

\bibitem[Fynbo {\etal}(2009a)]{FMH2009}
  Fynbo, J.~P.~U.,
  Malesani, D.,
  Hjorth, J.,
  Leitet, E.,
  Linne, S.,
  Ottosen, T.~A., \&
  Jakobsson, J.,
  2009a, GCNC 9140

\bibitem[Fynbo {\etal}(2009b)]{FJP2009}
  Fynbo, J.~P.~U.,
  Jakobsson, P.,
  Procheska, J.~X.,
  Malesani, D.,
  Ledoux, C.,
  de~Ugarte~Postigo, A.,
  Naradini, N.,
  Vreeswikjk, P.~M.,
  {\etal},
  2009b, \apjs, 525, 185

\bibitem[Gehrels {\etal}(2004)]{GCC2004}
  Gehrels, N.,
  Chincarini, G.,
  Ciommi, P.,
  Mason, K.~O.,
  Nousek, J.~A.,
  Wells, A.~A.,
  White, N.~E.,
  Barthelmy, S.~D.,
  {\etal},
  2004, \apj, 611, 1005

\bibitem[Genet {\etal}(2007)]{GDM2007}
  Genet, F.,
  Daigne, F., \&
  Mochkovitch, R.,
  2007, \mnras, 381, 732

\bibitem[Ghisellini {\etal}(2007)]{GGN2007}
  Ghisellini, G.,
  Ghirlanda, G.,
  Nava, L., \&
  Firmani, C.,
  2007, \apjl, 658, L75

\bibitem[Goad {\etal}(2007)]{GTB2007}
  Goad, M.~R.,
  Tyler, L.~G.,
  Beardmore, A.~P.,
  Evans, P.~A.,
  Rosen, S.~R.,
  Osborne, J.~P.,
  Starling, R.~L.~C.,
  Marshall, F.~E.,
  {\etal},
  2007, \aap, 476, 1401

\bibitem[Graham \& Driver(2005)]{GD2005}
  Graham, A.~W. \&
  Driver, S.~P.,
  2005, \pasa, 22, 118

\bibitem[Granot \& Sari(2002)]{GS2002}
  Granot, J. \&
  Sari, R.,
  2002, \apj, 568, 820

\bibitem[Guidorzi {\etal}(2009)]{GSM2009}
  Guidorzi, C.,
  Smith, R.,
  Mundell, C.~G.,
  Gomboc, A.,
  O'Brien, P., \&
  Tanvir, N.,
  2009, GCNC 9144

\bibitem[Hogg {\etal}(1997)]{HPM1997}
  Hogg, D.~W.,
  Pahre, M.~A.,
  McCarthy, J.~K.,
  Cohen, J.~G.,
  Blandford, R.,
  Smail, I., \&
  Soifer, B.~T.,
  1997, \mnras, 288, 404

\bibitem[Holland {\etal}(2007)]{HBG2007}
  Holland, S.~T.,
  Boyd, P.~T.,
  Gorosabel, J.,
  Hjorth, J.,
  Schady, P.,
  Thomsen, B.,
  Augusteijn, T.,
  Blustin, A.~J.,
  {\etal},
  2007, \aj, 133, 122

\bibitem[Jakobsson {\etal}(2004)]{JHF2004}
  Jakobsson, P.,
  Hjorth, J.,
  Fynbo, J.~P.~U.,
  Watson, D.,
  Pedersen, K.,
  Bj{\"o}rnsson, G., \&
  Gorosabel, J.,
  2004, \apjl, 617, L21

\bibitem[Jaunsen {\etal}(2008)]{JRW2008}
  Jaunsen, A.~O.,
  Rol, E.,
  Watson, D.~J.,
  Malesani, D.,
  Fynbo, J.~P.~U.,
  Milvang-Jensen, B.,
  Hjorth, J.,
  Vreeswijk, P.~M.,
  {\etal},
  2008, \apj, 681, 453

\bibitem[Kalberla {\etal}(2005)]{KBH2005}
  Kalberla, P.~M.~W.,
  Berton, W.~B.,
  Hartmann, D.,
  Arnal, E~.M.,
  Bajaja, E.,
  Morras, R.,
  P{\"o}ppel, W.~G.~L.,
  2005, \aap, 440, 775

\bibitem[Kann {\etal}(2006)]{KKZ2006}
  Kann, D.~A.,
  Klose, S., \&
  Zeh, A.,
  2006, \apj, 641, 993

\bibitem[Kouveliotou {\etal}(1993)]{KMF1993}
  Kouveliotou, C.,
  Meegan, C.~A.,
  Fishman, G.~J.,
  Bhat, N.~P.,
  Briggs, M.~S.,
  Koshut, T.~M.,
  Paciesas, W.~M., \&
  Pendleton, G.~N.,
  1993, \apjl, 413, 101

\bibitem[Kron (1980)]{K1980}
  Kron, R.~G.,
  1980, \apjs, 43, 305

\bibitem[Kumar {\etal}(2008)]{KNJ2008}
  Kuman, P.,
  Narayan, R., \&
  Johnson, J.~L.,
  2008, \mnras, 388, 1729

\bibitem[Le~Floc'h {\etal}(2002)]{LDM2002}
  Le~Floc'h, E.,
  Duc, P.-A.,
  Mirable, I.~F.,
  Sanders, D.~B.,
  Bosch, G.,
  Rodrigues, I.,
  Covrvoisier, T.~J.-L.,
  Mereghetti, S.,
  {\etal},
  2002, \apjl, 581, L81

\bibitem[Le~Floc'h {\etal}(2003)]{LDM2003}
  Le~Floc'h, E.,
  Duc, P.-A.,
  Mirable, I.~F.,
  Sanders, D.~B.,
  Bosch, G.,
  Diaz, R.~J.,
  Donzelli, C.~J.,
  Rodrigues, I.,
  {\etal},
  2003, \aap, 400, 499

\bibitem[Levan {\etal}(2006)]{LFR2006}
  Levan, A.,
  Fruchter, A.,
  Rhoads, J.,
  Mobasher, B.,
  Tanvir, N.,
  Gorosabel, J.,
  Rol, E.,
  Kouveliotou, C.,
  {\etal},
  2006, \apj, 647, 471

\bibitem[Liang {\etal}(2009)]{LLZ2009}
  Liang, E.-W.,
  L{\"u}, H.-J.,
  Zhang, B.-B., \&
  Zhang, B.,
  2009, \apj, 707, 328

\bibitem[Lilly {\etal}(1995)]{LTF1995}
  Lilly, S. J.,
  Tresse, L.,
  Hammer, F.,
  Crampton, D., \&
  Le~F{\`e}vre, O.,
  1995, \apj, 455, 108

\bibitem[Madau {\etal}(1998)]{MPD1998}
  Madau, P.,
  Pozzetti, L., \&
  Dickinson, M.,
  1998, \apj, 498, 106

\bibitem[Mangano {\etal}(2009)]{MBB2009}
  Mangano, V.,
  Barthelmy, S.~D.,
  Baumgartner, W.~H.,
  Burrows, D.~N.,
  Evans, P.~A.,
  Gehrels, N.,
  Guidorzi, C.,
  Holland, S~.T.,
  {\etal},
  2009, GCNC 9133

\bibitem[M{\'e}sz{\'a}ros \& Gruzinov(2000)]{MG2000}
  M{\'e}sz{\'a}ros, P., \&
  Grizinov, A.,
  2000, \apjl, 543, L35

\bibitem[Mizuno {\etal}(2001)]{MYM2001}
  Mizuno, N.,
  Yamaguchi, R.,
  Minzuno, A.,
  Rubio, M.,
  Abe, R.,
  Saito, H.,
  Onishi, T.,
  Yonekura, Y.,
  {\etal},
  2001, PASJ, 53, 971

\bibitem[Nousek {\etal}(2006)]{NKG2006}
  Nousek, J.~A.,
  Kouveliotou, C.,
  Grupe, D.,
  Page, K.~L.,
  Granot, J.,
  Ramirez-Ruiz, E.,
  Patel, S.~K.,
  Burrows, D.~N.,
  {\etal},
  2006, \apj, 642, 389

\bibitem[Panaitescu \& Kumar(2002)]{PK2002}
  Panaitescu, A., \&
  Kumar, P.,
  2002, \apj, 571, 779

\bibitem[Patat \& Piemonte(1998)]{PP1998}
  Patat, F., \&
  Piemonte, A.,
  1998, IAUC 6918

\bibitem[Pei(1992)]{P1992}
  Pei, Y.~C.,
  1992, \apj, 395, 130

\bibitem[Perley {\etal}(2009)]{PCB2009}
  Perley, D.~A.,
  Cenko, S.~B.,
  Bloom, J.~S.,
  Chen, H.-W.,
  Butler, N.~R.,
  Kocevski, D.,
  Prochaska, J.~X.,
  Brodwin, M.,
  {\etal},
  2009, \aj, 138, 1690

\bibitem[Piro {\etal}(2002)]{PFG2002}
  Piro, L.,
  Frail, D.~A.,
  Gorosabel, J.,
  Garmire, G.,
  Soffitta, P.,
  Amati, L.,
  Andersen, M.~I.,
  Antonelli, L.~A.,
  {\etal},
  2002, \apj, 577, 680

\bibitem[Poole {\etal}(2008)]{PBP2008}
  Poole, T.~S.,
  Breeveld, A.~A.,
  Page, M.~J.,
  Landsman, W.,
  Holland, S.~T.,
  Roming, P.~W.~A.,
  Kuin, N.~P.~M.,
  Brown, P.~J,
  {\etal},
  2008, \mnras, 383, 627

\bibitem[Predehl \& Schmitt(1995)]{PS1995}
  Predehl, P., \&
  Schmitt, J.~H.~M.~M.,
  1995, \aap, 293, 889

\bibitem[Rol {\etal}(2005)]{RWK2005}
  Rol, E.,
  Wijers, R.~A.~M.~J.,
  Kouveliotou, C.,
  Kaper, L., \&
  Kaneko, Y.,
  2005, \apj, 624, 868

\bibitem[Roming {\etal}(2005)]{RKM2005} 
  Roming, P.~W.~A.,
  Kennedy,T.~E.,
  Mason, K.~O.,
  Nousek, J.~A.,
  Ahr, L.,
  Bingham, R.~E.,
  Broos, P.~S.,
  Carter, M.~J.,
  {\etal},
  2005, Sp.\ Sci.\ Rev., 120, 95

\bibitem[Roming {\etal}(2006)]{RSF2006}
  Roming, P.~W.~A.,
  Schady, P.,
  Fox, D.~B.,
  Zhang, B.,
  Liang, E.,
  Mason, K.~O.,
  Rol, E.,
  Burrows, D.~N.,
  {\etal},
  2006, \apj, 652, 1416

\bibitem[Roming {\etal}(2009)]{RKO2009}
  Roming, P.~W.~A.,
  Koch, T.~S.,
  Oates, S.~R.,
  Porterfield, B.~L.,
  Vanden~Berk, D.~E.,
  Boyd, P.~T.,
  Holland, S.~T.,
  Hoversten, E.~A.,
  {\etal},
  2009, \apj, 690, 163

\bibitem[Salpeter(1955)]{S1955}
  Salpeter, E.~E.,
  1995, \apj, 121, 161

\bibitem[Salvaterra {\etal}(2009)]{SDC2009}
  Salvaterra, R.,
  Della~Valle, M.,
  Campana, S.,
  Chincarini, G.,
  Covino, S.,
  D'Avanzo, P.,
  Fernandez-Soto, A.,
  Guidorzi, F.,
  {\etal},
  2009, \nat, 461, 1258

\bibitem[Sari {\etal}(1999)]{SPH1999}
  Sari, R.,
  Piran, T., \&
  Halpern, J.~P.,
  1999, \apjl, 519, L17

\bibitem[Savaglio {\etal}(2009)]{SGL2009}
  Savaglio, S.,
  Glazebrook, K., \&
  Le~Borgne, D.,
  2009, \apj, 691, 182

\bibitem[Sbarufatti {\etal}(2009)]{SBB2009}
  Sbarufatti, B.,
  Barthelmy, S.~D.,
  Baumgartner, W.~H.,
  Burrows, D.~N.,
  Curran, P.~A.,
  Evans, P.~A.,
  Godet, O.,
  Guidorzi, C.,
  {\etal},
  2009, GCNC 9135

\bibitem[Scalo(1986)]{S1986}
  Scalo, J.~N.,
  1986, Fund.\ Cosm.\ Phys., 11, 1

\bibitem[Schady {\etal}(2007)]{SMP2007}
  Schady, P.,
  Mason, K.~O.,
  Page, M.~J.,
  De~Pasquale, M.,
  Morris, D.~C.,
  Romnano, P.,
  Roming, P.~W.~A.,
  Immler, S.,
  {\etal},
  2007, \mnras, 377, 273

\bibitem[Schady {\etal}(2009)]{SPO2009}
  Schady, P.,
  Page, M.~J.,
  Oates, S.~R.,
  Still, M.,
  De~Pasquale, M.,
  Dwelly, T.,
  Kuin, N.~P.~M.,
  Holland, S~.T.,
  {\etal},
  2009, \mnras, 401, 2773

\bibitem[Schlegel {\etal}(1998)]{SFD1998}
  Schlegel, D.~J.,
  Finkbeiner, D.~P., \&
  Davis, M.,
  1998, \apjs, 500, 525

\bibitem[S{\'e}rsic (1963)]{S1963}
  S{\'e}rsic, J.-L.
  1963, BAAA, 6, 41

\bibitem[S{\'e}rsic (1968)]{S1968}
  S{\'e}rsic, J.-L.
  1968, Atlas de Galaxias Australes, Cordoba: Observatorio Astronomico

\bibitem[Shao \& Dai(2007)]{SD2007}
  Shao, L., \&
  Dai, Z.~G.,
  2007, \apj, 660, 1319

\bibitem[Shao {\etal}(2008)]{SDM2008}
  Shao, L.,
  Dai, Z.~G., \&
  Mirabal, N.,
  2008, \apj, 675, 507

\bibitem[Shen {\etal}(2009)]{SWK2009}
  Shen, R.-F.,
  Willingale, R.,
  Kumar, P.,
  O'Brien, P.~T., \&
  Evans, P.~A.,
  2009, \mnras, 393, 598

\bibitem[Shen \& Zhang(2009)]{SZ2009}
  Shen, R.-F., \&
  Zhang, B.,
  2009, \mnras, 398, 1936

\bibitem[Strauss {\etal}(2002)]{SWL2002}
  Strauss, M.~A.,
  Weinberg, D.~H.,
  Lupton, R.~H.,
  Narayanan, V.~K.,
  Annis, J.,
  Bernardi, M.,
  Blanton, M.,
  Burles, S.,
  {\etal},
  2002, \aj, 124, 1810

\bibitem[Tanvir {\etal}(2008)]{TLR2008}
  Tanvir, N.~.R.,
  Levan, A.~J.,
  Rol, E.,
  Starling, R.~L.~C.,
  Gorosabel, J.,
  Priddey, R.~S.,
  Malesani, D.,
  Jakobsson, P.,
  {\etal},
  2008, \mnras, 388, 1743

\bibitem[Tanvir {\etal}(2009)]{TFL2009}
  Tanvir, N.~R.,
  Fox, D.~B.,
  Levan, A.~J.,
  Berger, E.,
  Wiersema, K.,
  Fynbo, J.~P.~U.,
  Cucchiara, A.,
  Kr{\"u}hler, T.,
  Gehrels, N.,
  {\etal},
  2009, \nat, 461, 1254

\bibitem[Trujillo {\etal}(2006)]{TFR2006}
  Trujillo, I.,
  F{\"o}rster~Schreiber, N.~M.,
  Rudnick, G.,
  Barden, G.,
  Franx, M.,
  Rix, H.-W.,
  Caldwell, J.~A.~R.,
  McIntosh, D.~H.,
  {\etal},
  2006 \apj, 650, 18

\bibitem[Uhm \& Beloborodov(2007)]{UB2007}
  Uhm, Z.~L., \&
  Beloborodov, A.~M.,
  2007, \apjl, 665, L93

\bibitem[Updike {\etal}(2009)]{UHG2009}
  Updike, A.~C.,
  Hartmann, D.~H.,
  Greiner, J., \&
  Klose, S.,
  2009, AIPC, 1133, 257

\bibitem[van~der~Horst {\etal}(2009)]{HKG2009}
  van der Horst, A.~J.,
  Kouveliotou, C.,
  Gehrels, N.,
  Rol, E.,
  Wijers, R.~A.~M.~J.,
  Cannizzo, J.~K., \&
  Racusin, J.,
  2010, \apjl, 711, L1

\bibitem[Wainwright {\etal}(2007)]{WBP2007}
  Wainwright, C.,
  Berger, E., \&
  Penprase, B.~E.,
  2007, \apj. 657, 367

\bibitem[Waxman \& Draine(2000)]{WD2000}
  Waxman, E., \&
  Draine, B.~T.,
  2000, \apj, 537, 796

\bibitem[Williams {\etal}(1990)]{WVT1990}
  Williams, P.~M.,
  van~der~Hucht, K.~A.,
  The, P.~S., \&
  Bouchet, P.,
  1990, \mnras, 247, 18

\bibitem[Wren {\etal}(2005)]{WVW2005}
  Wren, J.,
  Vestrand, W.~T.,
  Wozniak, P.,
  White, R., \&
  Evans, S.,
  2005, GCNC 3604

\bibitem[Yamazaki(2009)]{Y2009}
  Yamazaki, R.,
  2009, \apjl, 690, L118

\bibitem[Yasuda {\etal}(2007)]{YFS2007}
  Yasuda, N.,
  Fukugita, M., \&
  Schneider, D.~P.,
  2007, \aj, 134, 698

\bibitem[Zhang {\etal}(2006)]{ZFD2006}
  Zhang, B.,
  Fan, Y.~Z.,
  Dkys, J.,
  Kobayashi, S.,
  M{\'e}sz{\'a}ros, P.,
  Burrows, D.~N.,
  Nousek, J.~A., \&
  Gehrels, N.,
  2006, \apj, 642, 354

\end{thebibliography}
\end{document}